\documentclass[10pt,journal,compsoc]{IEEEtran}
\ifCLASSOPTIONcompsoc
  \usepackage[nocompress]{cite}
\else
  \usepackage{cite}
\fi
\ifCLASSINFOpdf
\else
\fi
\ifCLASSOPTIONcompsoc
  \usepackage[caption=false,font=footnotesize,labelfont=sf,textfont=sf]{subfig}
\else
  \usepackage[caption=false,font=footnotesize]{subfig}
\fi
\usepackage{fixltx2e}
\usepackage{dblfloatfix}

\usepackage{graphicx, verbatim,subfig}
\usepackage[T1]{fontenc}
\usepackage{algorithm, algorithmic,color}

\usepackage{amsmath, amsthm, amssymb}

\def\BibTeX{{\rm B\kern-.05em{\sc i\kern-.025em b}\kern-.08em
    T\kern-.1667em\lower.7ex\hbox{E}\kern-.125emX}}

\newcommand{\ktt}{\textcolor{black}}

\begin{document}

\title{Topology Abstraction Service for IP VPNs: Core Network Partitioning for Resource Sharing}
\author{Ravishankar~Ravindran,~\IEEEmembership{Senior Member,~IEEE,}
	   Changcheng~Huang,~\IEEEmembership{Senior Member,~IEEE,}
	   Krishnaiyan~Thulasiraman,~\IEEEmembership{Fellow,~IEEE,} 
	   and~Tachun~Lin,~\IEEEmembership{Member,~IEEE.} 
\IEEEcompsocitemizethanks{\IEEEcompsocthanksitem R. Ravindran is with Huawei Research Center, Santa Clara, CA, USA.\protect\\
E-mail: ravi.ravindran@huawei.com.
\IEEEcompsocthanksitem C. Huang is with Carleton University. K. Thulasiraman is with the University of Oklahoma.
T. Lin is with Bradley University.}}
\IEEEtitleabstractindextext{
\begin{abstract}
VPN service providers (VSP) and IP-VPN customers have traditionally maintained service demarcation boundaries between their routing and signaling entities. This has resulted in the VPNs viewing the VSP network as an opaque entity and therefore limiting any meaningful interaction between the VSP and the VPNs.  A key challenge is to expose each VPN to information about available network resources through an abstraction (TA)~\cite{ravi2006} which is both accurate and fair. In~\cite{ravi2013} we proposed three decentralized schemes assuming that all the border nodes performing the abstraction have access to the entire core network topology. This assumption likely leads to over- or under-subscription. In this paper we develop centralized schemes to partition the core network capacities, and assign each partition to a specific VPN for applying the decentralized abstraction schemes presented in~\cite{ravi2013}. First, we present two schemes based on the maximum concurrent flow and the maximum multicommodity flow (MMCF) formulations. We then propose approaches to address the fairness concerns that arise when MMCF formulation is used. We present results based on extensive simulations on several topologies, and provide a comparative evaluation of the different schemes in terms of abstraction efficiency, fairness to VPNs and call performance characteristics achieved.
\end{abstract}
}
\maketitle

\IEEEraisesectionheading{\section{Introduction}\label{sec:introduction}}
\IEEEPARstart{T}{opology} abstraction (TA) as a VPN service has been described in~\cite{ravi2006}. The objective of such a service is to provide the core topology information in an abstracted manner to the VPNs. The motivation for this service is the following:

\begin{itemize}
\item Providing TA information to the VPNs allows them to seek resources from the VSP with high degree of success.
\item In current provider networks, control plane scalability is a major concern. With TA service, as we will observe, the  VPNs achieve very good crankback ratio performance compared to VPNs that are not provided with any form of abstraction. This gain is significant as it reduces the call processing burden on the VSP which would otherwise be required to process these calls.
\end{itemize}

In~\cite{ravi2006} we established the above two points using simulation analysis. In~\cite{ravi2006} we also explained the notion of TA as applicable to VPNs, proposed a framework to realize it and proposed SLA parameters that could be used to provide service differentiation. In~\cite{ravi2013} we addressed a problem called VPN capacity sharing problem encountered in the context of TA service to VPNs. The objective of the problem was to generate abstractions so that the available resource is exposed in a fair manner to the VPNs subscribing to the TA service. We proposed solutions to this problem using decentralized schemes. We proposed three schemes in this context: maximum capacity scheme, mixed-bound scheme, and Steiner tree scheme. These schemes assume that all the border nodes performing the abstraction have access to the entire core topology, and generate TA to its hosted VPNs independent of one another. This form of TA generation leads to inefficient resource utilization and poor call performance. This suggests the need for a centralized approach  to combine all information and optimize resource allocation while maintaining fairness.

In this paper, we develop centralized schemes to partition the core network capacities, and assign each partition to a specific VPN for applying the decentralized abstraction schemes presented in~\cite{ravi2013}.  The main difference between the centralized form of TA generation considered in this paper   and the decentralized  TA generation process discussed in~\cite{ravi2013} is the use of a central server (CS) owned by the VSP to compute a resource partition subgraph for each VPN. These subgraph partitions are distributed to the border nodes, which use them to generate the desired TA that is then flooded to the VPN customer edge (CE) nodes. A key advantage of this process of TA generation is that it mitigates the problem of oversubscription encountered in the decentralized mode of TA generation. In this paper, our goal is to develop scalable centralized schemes that can allocate resources efficiently and maintain fairness among all the VPNs. Some preliminary results along these lines were presented in~\cite{ravi2007}.

The rest of the paper is organized as follows. \ktt{In Section~\ref{sec:notations} graph theory notations used in this paper are introduced. In Section \ref{sec:prevwork}, a summary of the essential ideas on topology abstraction described in [2] is presented. In Section~\ref{sec:vpncapacity} VPN core capacity sharing problem is stated.} 
In Section~\ref{sec:sec4}, partitioning schemes based on maximum concurrent flow (MConF) and the maximum multicommodity flow (MMCF) formulations are developed.  In Section~\ref{sec:sec5}, approaches to improve the fairness of the partitions generated by the MMCF method are developed.  We present in Section~\ref{sec:sec6} results of extensive simulations conducted on  several network topologies , and provide a comparative evaluation of the different schemes in terms of  abstraction efficiency, fairness to VPNs and call performance characteristics.

\section{Graph Theory Notations}
\label{sec:notations}
\begin{figure}[thpb]
\begin{center}
\includegraphics[keepaspectratio, scale=1]{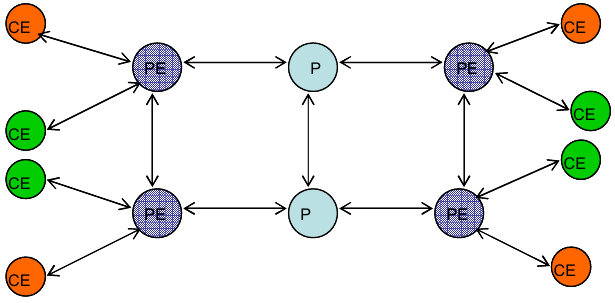} \caption{VSP Providing TA Service to two VPNs} \label{fig:1}
\end{center}
\vspace{-10px}
\end{figure}

We summarize in Table~\ref{tbl:1}
graph theory notations used in this paper. They are explained with reference to the network in Fig.~\ref{fig:1}.
\begin{table}[thbp]
  \centering
    \caption{Graph theory notations}\label{tbl:1}
    \footnotesize
    \begin{tabular}{|p{0.08\textwidth}|p{0.36\textwidth}|}
    \hline
    Graph Notations & Definition     \\\hline
    $G(V,E)$ & VSP core network     \\\hline
    $B$ & Set of all border (PE) nodes of the core network \\\hline
    $C_u,P_u$ & Set of CE/PE nodes of VPN $u$, here $P_u \subseteq B$ \\\hline
    $C_{u,b}$ & Set of VPNs hosted by border node $b$ \\\hline
    $G_{u,\ell}(V_u,E_u)$ & Abstract topology of VPN $u$ of topology type $\ell$ \\\hline
    $TS_u$ & TA SLA parameter set for VPN $u$ \\\hline
    $T_u/R_u$ & Abstract topology type subscription/Refresh interval of TA for VPN $u$ \\\hline
    $S(V_u,E_u)$ & Partition subgraph for VPN $u$ \\\hline
    $t_{ij}$ (or $t(e)$)  & Total capacity of edge $(i,j)$ or $e$ \\\hline
    $r_{ij}$ (or $r(e)$)  & Residual capacity of edge $(i,j)$ or $e$ \\\hline
    $U$ & Set of all VPNs provided TA service \\\hline
    $U_b$ & Set of VPNs hosted on node $b$ \\\hline
    $Z(x,y)$ & Set of VPNs common to border nodes $x$ and $y$ \\\hline
    $K$ & Set of all commodities in a multicommodity formulation \\\hline
    $K_v$ & Set of all source-destination border node pair commodity of VPN $v$ \\\hline
    $f_{mc}(k)$ & Flow of commodity $k$ corresponding to multicommodity flow (MMCF) formulation \\\hline
    $x^{u,k}_{i,j}$      & Resource partitioned on edge $(i,j)$ for VPN $u$ for source-destination commodity $k$ \\\hline
    $D(s_k,d_k)$ & Demand corresponding to source-destination commodity $k$ \\\hline
    \end{tabular}%
\end{table}%
As in Fig.~\ref{fig:1},
VSP's core topology is represented as a graph $G(V,E)$. Each directed link of the network $e_{i,j} \in E$ has a  capacity represented as $t_{i,j}$ (which we also denote as $t(e)$). This is the total capacity of the link. On the other hand, the capacity available on a link e at any given time is called residual capacity, denoted as $r_{i,j}$ (or $r(e)$). $B$ represents the set of all border PE nodes in graph $G(V,E)$. $U$ represents the set of all VPN customers subscribing to the TA service. Each border node $b \in B$ may support multiple VPN instances identified as $U_b$. For a VPN instance $u \in U$, we represent the sets of corresponding CE and PE nodes as $C_u$ and $P_u$ respectively. The set of CE nodes corresponding to a VPN instance $u \in U$ hosted on a border node $b$ is represented as $C_{u,b}$.  The '$P$' nodes in Fig.~\ref{fig:1}
are the provider nodes, which only participate in routing or switching the traffic.

\section{\ktt{Summary of Previous Work [2]}}\label{sec:prevwork}
\ktt{In this section we give a brief review of~\cite{ravi2013} where a detailed discussion of the topology abstraction (TA) service and decentralized topology abstraction algorithm are given.}

\subsection{\ktt{TA Service SLA Definition and Parameters:}}
\ktt{First, to enable a VSP to use TA service for generating service differentiation among the VPN customers, in~\cite{ravi2013} a new set of topology abstraction SLA (TA-SLA) parameters, which allows the VSP to customize the properties of the TA service to the requirements of the VPN is proposed. We next discuss the elements of the TA-SLA.}

\ktt{\textbf{Abstraction topology type parameter.} This parameter represents the type of abstract topology generated by the VSP for a VPN. The VSP uses this parameter to generate an abstract graph with a certain granularity before sharing it with the VPN. The optimizing objective for any form of TA is to minimize the complexity with respect to the granularity of the abstraction, while at the same time, maximizing the accuracy of the topology metric information that is being abstracted. Three forms of abstract topologies, namely source-star abstraction (SSA), star abstraction (SA), and simple node abstraction (SNA), which are also the most well-studied forms of abstractions in the context of hierarchical routing literature and used in~\cite{ravi2013}.}

\begin{figure}[htpb]
	\begin{center}
		\includegraphics[width=85mm]{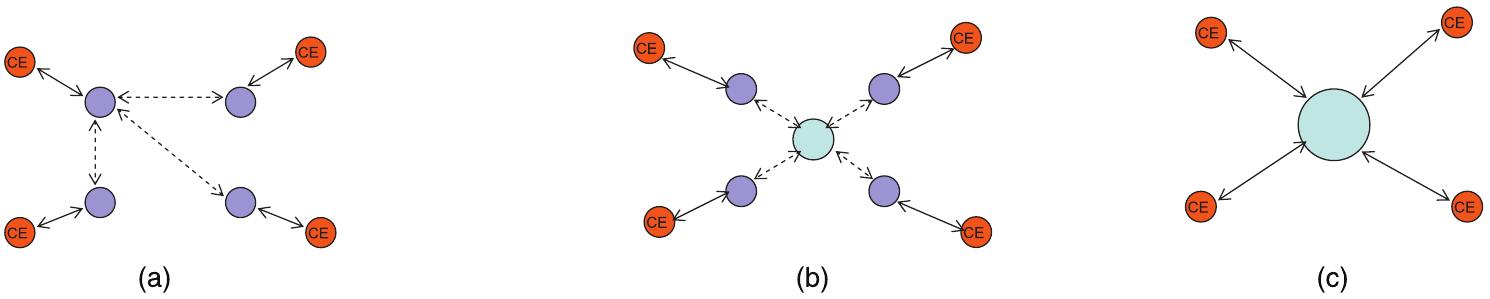} \caption{(a) source-star abstraction topology, (b) star abstraction topology, (c) simple node abstraction topology} \label{Fig2}
	\end{center}
\end{figure}

\ktt{Figs. \ref{Fig2}(a), \ref{Fig2}(b), and \ref{Fig2}(c) show these three types of abstract topologies. Among these the SSA type topology provides the finest granularity. Once a SSA type topology is generated, the other two can be generated through information aggregation. So we focus on the SSA type.}

The SSA type of abstraction, illustrated in Fig. \ref{Fig2}(a), has been adapted from~\cite{ravi2013}. In this case, a VPN is provided an abstraction that is a source- rooted tree. Here, the root of the tree is the PE node computing the abstraction for the corresponding VPN. The other nodes of the source-star topology are the PE nodes $P_k$ on which VPN of the same type are hosted. In the case where a CE node has adjacency with more than one PE node, multiple source-star abstractions would have to be provided to the CE node in order to take advantage of the routing diversity due to multiple PE connectivity. For the case of single CE-PE connectivity and for a given VPN $k$, the total message complexity of updating a CE is $O(|P_k| + |C_k|)$.

\ktt{\textbf{Abstract topology link metric.} This parameter identifies the choice of QoS metric information that is required to be associated with the virtual links of the abstract topology chosen as part of the abstraction topology type parameter. It is not necessary that all the VPNs subscribe the same set of QoS metric information. The choice of QoS metrics associated with the abstract topology could be determined by the purpose for which the TA service will be used for. Here, we use only one metric, namely available bandwidth. The virtual link metric associated with the nodes $x$ and $y$ for a VPN $k$ of the abstracted graph $G_{k,l}(V_k, E_k)$ is represented as $w_{k,l}(x,y)$.}

\ktt{\textbf{Abstract topology refresh interval metric.} The importance of this parameter is linked to the fact that the VSP provides periodic updates about resource availability by flooding the TA periodically in order to synchronize the core state information with that of the VPN's view of the core network. From this, we can see intuitively that the choice of this parameter decides the correctness of the TA information at any given point of time in relation to the link state of the VSP's core network. In a dynamic call request scenario, the ideal refresh rate (inverse of the refresh interval) desired by a VPN is expected to be a function of the mean arrival rate of bandwidth requests, dynamic nature of the state information in the core network, which will depend on factors like the load offered by all the VPNs and time of day, and the tolerance for the control overhead by the CE routers due to the frequency of update of abstraction information from the PE nodes. The abstract topology refresh rate negotiated by the VPN is also influenced by both the abstraction topology type and the abstract topology link metric chosen by the VPN.}

\begin{figure}[htpb]
	\begin{center}
		\includegraphics[height=50mm]{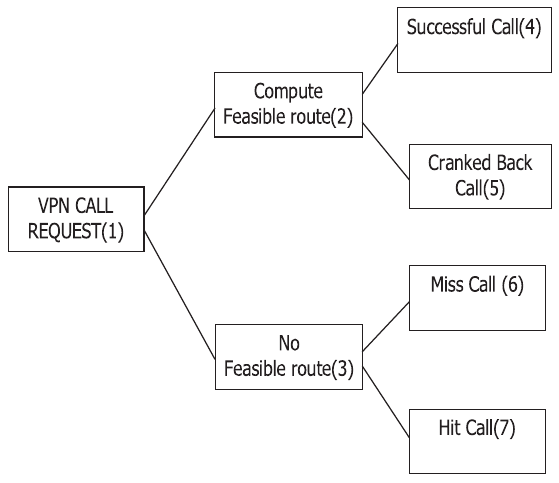} \caption{VPN call request scenarios} \label{Fig3}
	\end{center}
\end{figure}

\ktt{\textbf{Performance Metrics.} One of the significant benefits of abstracting VSP's core network and QoS information is to improve the call performance of the VPNs that require dynamic bandwidth service, while minimizing the overhead for sharing such information with the VPNs. Three call performance metrics have been defined to study the efficiency of the TA service, which are success, crankback, and misscall ratios. These metrics are explained with respect to Fig. \ref{Fig3} which shows the different possibilities of a bandwidth request from a VPN.}
\ktt{Starting from the root, denoted as (1), a VPN call request could have two possible outcomes. The first outcome is when a CE node finds that the requested capacity is not available and terminates the call locally. This is identified as (3). The second outcome is that the CE node finds that the requested capacity is available and sends the request to the corresponding PE node (2). For the case (3), we could have two possibilities. The first possibility is that the call has been wrongly terminated due to stale abstraction information, in which case we count it as a ``Miss'' call (6). The alternative possibility is that the call has been correctly terminated because of the VSP's inability to find the required path in the core network. We label such a call as a ``Hit'' call (7). For the case where a path is deemed to exist by the VPN, there can be two outcomes. The first is that the call does go through to the destination and the CE node receives a positive acknowledgement from the PE (4). The other is that the call crankbacks because of the inability of the PE node in the VSP to route the call (5). Crankbacked calls also waste signaling processing resources for the VSP. Success ratio measures the percentage of correct decisions in terms of both successfully computing a path (with successful signaling) and rejecting the call locally using the abstraction provided by the VSP. Crankback ratio measures the percentage of calls that resulted in successful path computation locally but is cranked back by the VSP. Misscall ratio measures the proportion of calls which were terminated wrongly at the VPN's end, in spite of sufficient resources in the core network to satisfy the call. For each VPN, success ratio is to be maximized while crankback and miscall ratio are to be minimized. These metrics are summarized below.}

\ktt{\textbf{Success ratio.} The success ratio is a measure of a VPN making a right routing decision using the abstraction provided to it by the VSP. This includes successful calls and the hit calls. The availability of the requested capacity is verified by recomputing the path with the exact state of the network}

{\small
\text{Success Ratio} = 
\[
\frac{\text{Number of calls (correctly accepted + calls correctly rejected)}}{\text{Total number of calls}}
\]
}

\ktt{\textbf{Crankback ratio.} A call would crank back, if there were no feasible path with the requested capacity in the core. The crankback ratio is defined as the ratio of the number of calls that have been cranked back to the total number of path requests made by the VPN}
\[\text{CrankBack Ratio} =
\;\;\;\; \frac{\text{Number of calls cranked back}}{\text{Total number of calls}}
\]

\ktt{\textbf{Misscall ratio.} The misscall ratio is the ratio of calls that have been wrongly terminated locally at the VPN's end (even though there is enough resource to accommodate the call) to the total number of calls originated by the VPN; ideally, a successful TA service implementation should have a miss call ratio of zero}
\[
	\text{MissCall Ratio} = 
		\frac{\text{Number of wrongly rejected calls}}{\text{Total number of calls}}
\]
\ktt{\textbf{Average network utilization.} This metric refers to the ratio of total link capacity utilized by active VPN bandwidth requests (aggregate utilized link capacity) to the total link capacity (aggregate link capacity)}
\text{Average Network Utilization} =
\[
\;\;\;\; \frac{\text{Aggregate utilized link capacity}}{\text{Aggregate link capacity}}
\]

\ktt{To generate the TA for the VPNs, we assume the availability of the entire core network topology, locally in a border node. In addition, the algorithms also require VPN specific information, which would include information about the VPNs hosted on each border node, along with the residual capacity of the VPN access link connecting the PE and the CE nodes. The core topology information is readily available as part of link state protocols, such as OSPF or ISIS, which are the protocols of choice in most VSP networks. The VPN specific information required by the abstraction schemes can be obtained from a TA service specific database populated by appropriate protocol extensions that include VPN membership information and the associated TA-SLA parameters discussed earlier. This in- formation can be either locally available at the PE node or accessed centrally from a centralized database.}

\ktt{\textbf{Fairness:} An important factor to consider during the abstraction process is the TA-SLA parameters, particularly the abstraction topology type parameter to offer service differentiation. To enable this service differentiation, we assume that the VSP adopts a model where the goal is to always expose the available capacity differently to different types of TA while exposing the available capacity equally among all the VPNs that have subscribed to the same abstract topology type parameter $T_k$. This fairness policy is applied locally by each border node $b$, while generating abstract topologies for the set of VPNs $U_b$ hosted by it. The criterion we adopt can be stated as follows: For a pair of VPNs $(i, j)\in U_b$ with the same abstract topology type parameter $l$, with abstract graphs $G_{i,l}(V_i, E_i)$ and $G_{i,l}(V_j, E_j)$ and with a common pair of nodes $x$ and $y$, the VSP imposes the rule that the exposed capacity associated with virtual link connecting the nodes $x$ and $y$ should be equal, that is $w_{i,l}(x,y) = w_{j,l}(x,y)$. This ensures that two VPNs with the same TA type will be exposed the same amount of bandwidth for each virtual link connecting the nodes $x$ and $y$ in the abstract topology.}

\ktt{\textbf{Topology Abstraction Problem and Abstraction Algorithms:}}
\ktt{The challenge faced during the abstract topology generation process is the fact that the process of generating TAs may lead to oversubscription of the available core resource, which could result in poor VPN call performance and high signaling cost for VSPs under high load conditions. Oversubscription results because of two reasons: first due to the decentralized approach of TA generation wherein each border node generates TAs for the VPNs without coordinating with one another; second, the degree of oversubscription also depends on the abstraction algorithm that is used to generate the abstraction for each VPN by a border node. The VPN-TA problem stated next is to address the problem of oversubscription arising due to the second factor in a decentralized context of TA generation. The objectives of this problem are the following:}
\ktt{
\begin{itemize}
	\item Provide the VPNs with an accurate representation of available capacity considering the case of having to satisfy simultaneous VPN calls during high load condition. This objective also correlates with the objective of maximizing the call performance of the VPNs.
	\item Maximize utilization of the VSP core network.
	\item Generate fair abstraction which aligns with the fairness policy presented earlier.
\end{itemize}}

\ktt{\textbf{VPN topology abstraction (VPN-TA) problem.} Given a set of VPNs $U_b$ hosted on the border node $b$, each VPN instance $i \in U_b$ is to be provided with an abstract topology $G_{i,l}(V_i, E_i)$. The objective is to device a methodology to allocate virtual capacities to the links in $E_i$ so that the VSP maximizes the probability of each VPN making a correct decision of successfully computing or rejecting a path locally in the context of the TA service.}

\ktt{In~\cite{ravi2013} three algorithms for the above problem are proposed. 
All the three abstraction schemes have been proposed to maximize the call success ratio. The three schemes vary in their nature from being aggressive or conservative in terms of associating virtual capacity to the link in the abstract topology. The nature of these algorithms leads to different performance results varying in tradeoffs between the three call performance metrics, i.e., the success, crankback, and misscall ratio.}
\section{VPN Capacity Sharing Problem}
\label{sec:vpncapacity}
As part of the TA service, each VPN is served with an abstract topology of type $l$. We discussed different types of abstract topologies in~\cite{ravi2006}; they are source-star (SSA), star, and  simple node abstract topologies. The topology of the latter two types can be generated from the first type through an information aggregation process which results in information loss and therefore differentiated services as discussed in~\cite{ravi2006}. In this paper we discuss TA generation with respect to SSA, the difference in performance when star or simple node forms of abstraction is chosen by a VPN should be similar to that discussed in~\cite{ravi2006}. Different from~\cite{ravi2006}, the abstract topology for a VPN $u$ will be derived from a partition subgraph denoted as $S(V_u, E_u)$.  Centralized schemes to generate these partition subgraphs are the main focus of this paper. With this in view, we now define the VPN core capacity sharing problem. We start with a restatement of the objectives.

First, the goal of a TA service is to enable better call performance to the VPNs making dynamic capacity requests; in other words, the goal is to maximize call performance.  We use the VPN success ratio as a parameter to measure VPN call performance. This parameter is a measure of making right bandwidth request decisions by a VPN using the abstraction provided to it by the VSP. This includes the calls that are computed and signaled successfully by the CE node as well those rejected correctly by the VPN locally. The local call rejections happen due to insufficient resources to accommodate the VPN call request because of increased load conditions.

The second objective is to ensure the best possible use of the VSP's available core capacity resource. Hence, the objective is also to maximize a VSP's network utilization. This also correlates with the goal of maximizing the revenue generated out of the TA service.

The third criterion we consider while computing partition subgraphs for the VPNs is to ensure that the schemes are fair to all the VPNs. This fairness policy requires that we expose the available capacity equally to all the VPNs that have the same abstract topology type $T_k$, while maintaining the desired property that different abstract topology types result in different service levels.
Considering these objectives, we define the VPN core capacity sharing problem.

\textbf{VPN Core Capacity Sharing Problem (VPN-CS):}\label{sec:sec3}\\
Given a graph $G(V,E)$ representing the VSP's core network that provides topology abstraction service to the set of VPNs, $U$, the objective is to compute fair  partitions  $S(V_u,E_u)$ for each VPN $u\in U$ of the network so as  to maximize the VPN call success ratio and core network utilization. If  $x_e^u$ represents the resource identified as part of subgraph $S(V_u,E_u)$ on edge $e\in E$ and for  VPN $u\in U$,  then $\displaystyle\sum_{u\in U}x_e^u \le t(e) ,\ \forall e\in E$ must be satisfied.

\section{Multicommodity Flow Formulation of the VPN-CS Problem}
\label{sec:sec4}
In view of the relationship between the maximization objectives of the VPN-CS problem and the total abstracted capacity considered during the partitioning phase (that is, the sum of the link capacities of the partition graphs $S(V_u,E_u)$ for each $u\in U$),  we first  present a  multicommodity flow formulation that maximizes the total abstracted capacity. The fairness criterion discussed in Section~\ref{sec:prevwork} is also incorporated in this formulation.  	
In this formulation, $K_v$ represents the set of all source-destination pairs of VPN $v \in U$ derived from the set $P_v$. We define the set of commodities $K$ as the set of all the VPN commodities, i.e. $K= \{K_v | \forall v \in U\}$. Let the net flow achieved for each source-destination commodity $k \in K_v$ of VPN $v$ be $f_{v,k}$. The variable $x^{v,k}_{i,j}$  denotes the logical resource assigned to source-destination pair $(s_k,d_k)\in K_v$ on an edge $(i,j)\in E$. Using these definitions, the VPN-CS problem can be formulated as in (1)-(5) with the objective of maximizing the aggregate resources considered as part of the subgraph partitions.
\begin{center}{\bf Formulation for VPN Core Capacity Sharing Problem}\end{center}
{\footnotesize
\begin{align}
&\nonumber\text{Maximize}  \displaystyle\sum_{v\in U}\displaystyle\sum_{k\in K_v}\displaystyle\sum_{(i,j)\in E}\ktt{x}_{i,j}^{v,k} \\
&s.t. \displaystyle\sum_{(i,j)\in E}\ktt{x}_{i,j}^{v,k} - \displaystyle\sum_{(j,i)\in E}\ktt{x}_{j,i}^{v,k}  =  f_{v,k}\ , \forall v\in U, k\in K_v, i=s_k \label{eq:f1_1}\\
&\displaystyle\sum_{(i,j)\in E}\ktt{x}_{i,j}^{v,k} - \displaystyle\sum_{(j,i)\in E}\ktt{x}_{j,i}^{v,k}  =  0, \ \ \forall v\in U, k\in K_v, i \ne s_k, i\ne d_k \label{eq:f1_2}\\
&\displaystyle\sum_{(i,j)\in E}\ktt{x}_{i,j}^{v,k} - \displaystyle\sum_{(j,i)\in E}\ktt{x}_{j,i}^{v,k}  =  -f_{v,k}\ , \ \ \forall v\in U, k\in K_v, i=d_k \label{eq:f1_3}\\
&\displaystyle\sum_{v\in U}\displaystyle\sum_{k\in K_v}\ktt{x}_{i,j}^{v,k} \le t_{i,j}, \ \ \forall (i,j) \in E \label{eq:f1_4}\\
&f_{v_1, k} - f_{v_2, k} = 0, \ \ \forall s_k, d_k \in B, \forall (v_1, v_2) \in Z(s_k, d_k) \label{eq:f1_5} \\
&\nonumber x_{i,j}^{v,k} \ge 0, \phantom{00} f_{v,k} \ge 0
\end{align}
\vspace{-10px}
}

In the formulation, \eqref{eq:f1_1}--\eqref{eq:f1_3} enforce the supply-demand conservation condition for each VPN commodity.  \eqref{eq:f1_4} enforces the capacity constraint for each edge of the core graph.  \eqref{eq:f1_5} enforces the fairness constraint during the subgraph partition computation process with the goal of sharing resources equally among all the VPNs. We define $Z(s,d)$ as the set of VPNs having a common source-destination border node pair $(s,d)$.  The constraint  \eqref{eq:f1_5} ensures equal sharing of resources by enforcing the aggregate flow, i.e, $f_{v_1,k}, f_{v_2, k}$ for any two VPNs $v_1,v_2 \in Z(s_k,d_k)$ to be equal. The optimum flow values $x^{v,k}_{i,j}$ define the capacity of the link $(i,j)$ in the partition graph  for each $v\in U$ and commodity $k$.  The worst case scenario of the formulation can be assessed assuming that all the VPNs in $U$ are hosted on all the border nodes in $B$. In this case, the number of variables in the formulation is $O(|E|\times|U|\times|B|^2 + |U|\times|B|^2)$, and the number of possible constraints would be in the order  of  $O(|U|\times|B|^2\times|V| + |U|^2\times|B|^2)$. Since the goal of the TA service is to enable dynamic bandwidth requests based on a topology abstraction in smaller time scales, solving the problem to optimality for large numbers of variables and constraints as in the above formulation using linear programming tools will not be efficient. So, we propose solutions based on a variant of the multicommodity flow problem, namely, the maximum concurrent flow (MConF) problem, whose objective is to maximize the aggregate commodity flow, while ensuring fairness among commodity flows.
\subsection{Maximum Concurrent Flow Formulation of the VPN-CS Problem}
\label{sec:mcf}
The MConF problem is defined as follows. Given a network $G(V,E)$  and a set $K$  of source-destination pairs $(s_i, d_i), i=1,2, \ldots |K|$, assume that demands $D(s_i, d_i), \forall i=1,2, \ldots |K|$ are known. The objective of the MConF problem is to maximize the factor $\beta$ such that there exists a flow that satisfies the demand $\beta \times D(s_i, d_i), \forall i= 1, 2, \ldots , k$. The node-link formulation of the MConF problem is as follows.

{\bf Maximum Concurrent Flow (MConF) Formulation:}
{\footnotesize
\begin{align}
&\nonumber\text{Maximize} \ \  \beta \\
&s.t. \displaystyle\sum_{(i,j)\in E}\ktt{x}_{i,j}^{k} - \displaystyle\sum_{(j,i)\in E}\ktt{x}_{j,i}^{k}  =  \beta \times D(s_k, d_k),  \ \ \forall \ \  k\in K_v, i=s_k \label{eq:f2_1}\\
&\displaystyle\sum_{(i,j)\in E}\ktt{x}_{i,j}^{k} - \displaystyle\sum_{(j,i)\in E}\ktt{x}_{j,i}^{k}  =  0,  \ \ \forall \ \  k\in K_v, i \ne s_k, i\ne d_k \label{eq:f2_2}
\end{align}
\begin{align}
&\displaystyle\sum_{(i,j)\in E}\ktt{x}_{i,j}^{k} - \displaystyle\sum_{(j,i)\in E}\ktt{x}_{j,i}^{k}  =  -\beta \times D(s_k, d_k),   \ \ \forall \ \  k\in K_v, i=d_k  \label{eq:f2_3}\\
&\displaystyle\sum_{k\in K}\ktt{x}_{i,j}^{k} \le t_{i,j},  \ \ \forall \ \  (i,j) \in E \label{eq:f2_4} \\
&\nonumber x_{i,j}^{k} \ge 0, \phantom{00}  \beta \ge 0
\end{align}
}
The value of $\beta$ is called the throughput of the maximum concurrent flow formulation. The objective of the MConF problem enforces fairness by satisfying the same fraction of the demand for all the commodities, which is also in line with the fairness policy discussed in Section \ref{sec:prevwork}.

With regard to the solution for the MConF problem for online implementation,~\cite{farhad1990} proposed the first approximation algorithm for the MConF problem for networks with equal demands and undirected edges. Several researchers have since proposed enhancements to this algorithm in order to improve its complexity on the generalized version of the problem with arbitrary capacities. Later, Garg and Konemann~\cite{garg1998} proposed a simple approximation algorithm to the MConF problem. Fleischer~\cite{lisa1999} presented improvements to the maximum concurrent flow approximation algorithm proposed in~\cite{garg1998}, with better running time complexity, particularly, when the graph is sparse and has a larger number of commodities satisfying the condition $|K| > |E|/|V|$.

Applying MConF to VPN-CS problem is not straight forward for two reasons. First, the commodity demands $D(s_k,d_k)$ for the VPN-CS problem are not known a priori. Secondly, if the commodity is defined for each VPN $v \in U$ and source-destination commodity $k \in K_v$, the complexity of the number of commodities would be $O(|U|\times|B|^2)$, which may not be scalable for online implementation.

We address these two concerns as follows. Since one of the objectives of VPN-CS problem is to maximize the link resources considered during the partitioning phase, we initialize $D(s_k,d_k)$ to $\alpha(k)$, where $\alpha(k)$ is the maximum flow  between the source-destination pair $(s_k,d_k)$. In order to reduce the complexity of the execution time and memory requirement, we propose a two step optimization approach. The first step is to apply the MConF formulation on an aggregated set of source-destination pairs. For this scheme, we define a commodity $k$ in the aggregated version as source-destination border node pair $(s,d)$ that has a potential VPN bandwidth request from $s$ to $d$. This is the case if the source-destination border nodes $s$ and $d$ host at least one common VPN; in this case, it will satisfy the criteria     $ |(U_s \cap U_d)| \ge 1$. This aggregation of commodities reduces commodity complexity to $O(|B|^2)$. With this simplification, we propose a centralized heuristic to be executed by the CS in order to solve the VPN-CS problem as shown in the pseudo code in Algorithm 1.

\begin{algorithm}[thpb]
{\small
\begin{algorithmic}[1]
\REQUIRE{ $G = ( V, E)$, $B$, $U$, $P_k\ \ \forall\ k \in U$}
\ENSURE{$S(V_v, E_v), v \in U$}
\FORALL{potential pair $(s, d), s, d \in B$}
	\STATE{Identify the set of VPNs $Z(s, d)$ and define commodity set $K = \{(s,d): | Z(s,d) | \ge 1\}$}
\ENDFOR
\FORALL{commodity pair $(s_k, d_k), \forall k \in K$}
	\STATE{Compute the maxflow $\alpha(k)$}
\ENDFOR
\FORALL{boarder node pair $(s_k, d_k), k\in K$}
	\STATE{Set the supply for $s_k$ to $\alpha(k)$}
	\STATE{Set the demand at $d_k$ to $- \alpha(k)$}
\ENDFOR
\STATE{Solve the maximum concurrent flow problem for commodity set $K$ with respect to graph $G(V, E)$}
\COMMENT{Note: The solution represents the resource associated with commodity $k$ over edge $(i,j) \in E$}
\FORALL{commodity pair $x_{i,j}^k, (s_k, d_k) \in K, (i, j) \in E$}
	\IF{$x_{i,j}^k > 0$}
		\STATE{$x_{i,j}^{k,v} = \frac{x_{i,j}^k}{Z(s_k, d_k)}$}
		\COMMENT{Note: we divide the edge flows fairly among the VPNs in the set $Z(s_k,d_k)$. Here, $x_{i,j}^{k,v}$ is the capacity assigned to VPN $v$ on link $(i,j)$ for VPN source-destination pair $(s_k,,d_k ) \in K_v$ }
	\ENDIF
\ENDFOR
\FORALL{VPN $v \in U$}
	\STATE{Create partition subgraph $S(V_v, E_v)$ where $V_v \subset V$ and $E_v \subset E$ are the set of all links such that for at least one commodity $k \in K_v$ of VPN $v$, $x_{i,j}^{k,v} > 0$. The capacity of each link $(i,j)$ in $S(V_v, E_v)$ is the sum of the $x_{i,j}^{k,v}$ flows for all commodities $k \in K_v$}
\ENDFOR
\FORALL{VPN $v$}
	\STATE{Distribute $S(V_v, E_v)$ to the border nodes in $P_v$}
\ENDFOR
\caption{MConF based partitioning heuristic for VPN core capacity sharing problem}
\label{alg:mcfbps}
\end{algorithmic}
}
\end{algorithm}

Based on our previous discussion, Lines 1 -- 11
are self-explanatory. Lines 12--16
iterate over each commodity $k \in K$. For each commodity $k$ and the set of VPNs $v \in Z(s_k,d_k)$ sharing the source-destination pair $(s_k,d_k)$, we set the logical capacity $x_{i,j}^{v,k}$ over link $(i,j)$ for VPN $v \in Z(s_k,d_k)$ equal to $x_{i,j}^{k} / |Z(s_k,d_k)|$; this means we divide  the MConF edge flow $x_{i,j}^{k}$ equally among all the VPNs $v \in Z(s_k,d_k)$.
Lines 12--16 are
inline with our policy of sharing link resources fairly among  the set of VPNs $Z(s_k,d_k)$ whose flows share a link; another step of fairness is applied by the border nodes, as per~\cite{ravi2006}, when the final abstractions are generated.
Lines 17--19
use the capacities logically partitioned for each of the VPN $v \in U$ on each link $(i,j) \in E$ in order to realize the VPN subgraph $S(V_v, E_v)$, which is obtained by constructing a subgraph with positive edge flows assigned for VPN $v$. Once $S(V_v, E_v)$ is determined, it is then distributed to the nodes in $P_v$ (those border nodes hosting VPN $v$ ) as in Lines 20--22.
The partition subgraphs will be used by the PE nodes to generate TAs using the abstraction schemes in~\cite{ravi2013}.

For online implementation,~\cite{lisa1999} proposes the fastest $\epsilon$-approximation algorithm for the MConF problem; this can be applied in Line 11
of the MConF based heuristic. With respect to the complexity of the MConF based partitioning scheme, it is dominated by Lines 4--6 and  11
, which results in the overall complexity of   $O(\epsilon^2|E|\times(|E|+|K| )+|K|\times|V|^3)$.
\subsection{Multicommodity Flow Based Formulation of the VPN-CS Problem}
\label{sec:mcfvpn}
\begin{figure}[htpb]
\begin{center}
\includegraphics[height=30mm]{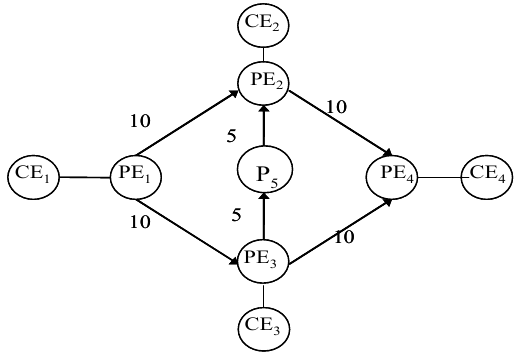} \caption{Example demonstrating MConF drawback} \label{fig:2}
\end{center}
\end{figure}
We now draw attention to a drawback of the MConF based partitioning scheme and propose an approach based on the maximum multicommodity flow (MMCF) theory in order to address it. Figure~\ref{fig:2} shows a VSP core network. Let us consider three commodities $k(1)={PE_1,PE_4}, k(2)={PE_1,PE_2}$, and $k(3)={PE_3,PE_2}$.

The throughput achieved by the MConF formulation with the values of the commodities set to their maximum flow values of $\{20, 15, 5\}$ is $\beta = 0.5$, and the aggregate commodity flow (which is the sum of flows between the three source-destination pairs) is 20 units. This happens because the edges of the paths corresponding to a commodity's maximum flow are also shared by other commodities, which results in $\beta < 1$ forcing the same fraction of the demand to be satisfied for all the commodities. Through observation, we can see that the aggregate commodity flow, which can be achieved when the flows of each commodity are increased independently, is 25 units. This inefficiency can be addressed by maximizing the sum of the independent commodity flows; in other words, we use the objective of maximizing $\displaystyle\sum_k{\beta_k \times D(s_k, d_k)}$. The modified MconF (M-MConF) problem is stated next.
\begin{center}{\bf Modified Maximum Concurrent Flow (M-MConF) Formulation:}\end{center}
{\footnotesize
\begin{align}
&\nonumber\text{Maximize} \ \  \displaystyle\sum_k{\beta_k \times D(s_k, d_k)} \\
&\nonumber\text{Subject to:} \phantom{ \displaystyle\sum_{v_i\in U}\displaystyle\sum_{k\in K}\displaystyle\sum_{(i,j)\in E}x_{i,j}^{v,k}} \\
&\displaystyle\sum_{(i,j)\in E}x_{i,j}^{k} - \displaystyle\sum_{(j,i)\in E}x_{j,i}^{k}   =  \beta_k \times D(s_k, d_k),  \ \ \forall k\in K, i=s_k \label{eq:f3_1}\\
&\displaystyle\sum_{(i,j)\in E}x_{i,j}^{k} - \displaystyle\sum_{(j,i)\in E}x_{j,i}^{k}    =  0,  \ \ \forall k\in K, i \ne s_k, i\ne d_k \label{eq:f3_2}\\
&\displaystyle\sum_{(i,j)\in E}x_{i,j}^{k} - \displaystyle\sum_{(j,i)\in E}x_{j,i}^{k}    =  -\beta_k \times D(s_k, d_k),  \forall \ \  k\in K, i=d_k  \label{eq:f3_3}\\
&\displaystyle\sum_{k\in K}x_{i,j}^{k} \le t_{i,j},  \ \ \forall  (i,j) \in E \label{eq:f3_4} \\
&\nonumber x_{i,j}^{k} \ge 0, \phantom{00}  \beta \ge 0
\end{align}
}
Replacing $\beta_k \times D(s_k,d_k)$ by $f_{mc}(k)$  in the MMConF formulation we get the following equivalent maximum multicommodity flow formulation whose objective is to maximize the total flow for all the commodities. The node-link MMCF formulation is as follows:

\begin{center}{\bf Maximum Multicommodity Flow (MMCF) Formulation:}\end{center}
{\footnotesize
\begin{align}
&\nonumber\text{Maximize} \ \  \displaystyle\sum_{k\in K}{f_{mc}(k)} \\
&\nonumber\text{Subject to:} \phantom{ \displaystyle\sum_{v_i\in U}\displaystyle\sum_{k\in K}\displaystyle\sum_{(i,j)\in E}x_{i,j}^{v,k}}\\
&\displaystyle\sum_{(i,j)\in E}x_{i,j}^{k} - \displaystyle\sum_{(j,i)\in E}x_{j,i}^{k}   =  f_{mc}(k),  \ \ \forall  k\in K, i=s_k \label{eq:f4_1}\\
&\displaystyle\sum_{(i,j)\in E}x_{i,j}^{k} - \displaystyle\sum_{(j,i)\in E}x_{j,i}^{k}   =  0,  \ \ \forall  k\in K, i \ne s_k, i\ne d_k \label{eq:f4_2}
\end{align}
\begin{align}
&\displaystyle\sum_{(i,j)\in E}x_{i,j}^{k} - \displaystyle\sum_{(j,i)\in E}x_{j,i}^{k}   =  -f_{mc}(k),  \ \ \forall  k\in K, i=d_k  \label{eq:f4_3}\\
&\displaystyle\sum_{k\in K}x_{i,j}^{k} \le t_{i,j},  \ \ \forall (i,j) \in E \label{eq:f4_4} \\
&\nonumber x_{i,j}^{k} \ge 0, \phantom{00}  \beta \ge 0
\end{align}
}
Here, $f_{mc}(k)$ is the aggregate MMCF flow variable corresponding to commodity $k$. The objective of MMCF is to maximize the total commodity flow. As in previous formulations, (14) -- (16) represent the supply-demand constraints and (17) the capacity bound for edge $(i,j)\in E$.
Next, we propose in Algorithm 2
a MMCF based heuristic to solve the VPN-CS problem.

\begin{algorithm}[tph]
{\small
\begin{algorithmic}[1]
\REQUIRE{ $G = ( V, E)$, $B$, $U$, $P_k\ \ \forall\ k \in U$}
\ENSURE{$S(V_v, E_v), v \in U$}
\FORALL{potential border node pair $(s, d), s, d \in B$}
	\STATE{Identify the set $Z(s, d)$ and define commodity set \\$K = \{(s,d): | Z(s,d) | > 0\}$}
\ENDFOR
\STATE{Solve the maximum multicommodity flow problem for commodity set $K$ with respect to graph $G(V, E)$}
\COMMENT{Note: The solution $x_{i,j}^k$ represents the resource associated with commodity $k$ over edge $(i,j) \in E$}
\FORALL{commodity pair $x_{i,j}^k, (s_k, d_k) \in K, (i, j) \in E$}
	\IF{$x_{i,j}^k > 0$}
		\STATE{$x_{i,j}^{k,v} = \frac{x_{i,j}^k}{Z(s_k, d_k)}$}
		\COMMENT{Note: we divide the edge flows $x_{i,j}^k$ fairly among the VPNs in the set $Z(s_k,d_k)$. Here, $x_{i,j}^{k,v}$ is the capacity assigned to VPN $v$ on link $(i,j)$ for VPN source-destination pair $(s_k,,d_k ) \in K_v$}
	\ENDIF
\ENDFOR
\FORALL{VPN $v \in U$}
	\STATE{Create partition subgraph $S(V_v, E_v)$ where $V_v \subset V$ and $E_v \subset E$ are the set of all links such that $x_{i,j}^{k,v} > 0$ for at least one commodity $k \in K_v$ of VPN $v$. The capacity of each link $(i,j)$ in $S(V_v, E_v)$ is the sum of the $x_{i,j}^{k,v}$ flows for all commodities $k \in K_v$}
\ENDFOR
\FORALL{VPN $v$}
	\STATE{Distribute $S(V_v, E_v)$ to the border nodes in $P_v$}
\ENDFOR
\caption{MMCF based partitioning heuristic for VPN core capacity sharing problem}
\label{alg:mmfbps}
\end{algorithmic}
}
\end{algorithm}

As in the case of the MConF based partition generation scheme, MMCF begins in Lines 1--3
by initializing the set $Z(s,d)$ for all potential source-destination pairs $(s,d)\in B$; this is then used to determine the commodity set $K$.  Line 4
solves the MMCF problem. For an online implementation, the $\epsilon$-approximation algorithm from~\cite{lisa1999} can be applied. This results in partitions associated with aggregate source-destination commodity   $x_{i,j}^k \forall k\in K$ and $(i,j)\in E$. Lines 5--15
are similar to the fair partitioning process discussed with respect to the MConF based partitioning solution. The MMCF heuristic is expected to achieve better network utilization compared to MConF heuristic; this is due to an increase in aggregate edge capacity considered during the partitioning process. For an online implementation, the overall complexity of this heuristic is dominated by Line 4
, where we apply the $\epsilon$-approximation algorithm presented in~\cite{lisa1999}, which results in a complexity of $O(\epsilon^2|E|(|E|+ |V|\times \log|E|)\log(|V|))$.
\section{Improving Fairness of MMCF Partitioning Scheme}
\label{sec:sec5}

Though the MMCF based partitioning scheme maximizes the aggregate flow, it may not achieve the desired objective of maximizing $\beta_k$ for each commodity $k$ in a fair manner. Hence, the final $\beta_k$ values may end up being very optimistic for a few commodities and conservative for others, which leads to unfair resource partitioning among the commodities. Since the goal of maximizing the VPN success ratio also requires  partitioning the network capacity as fairly as possible among all the VPNs, we define another problem called the fair partitioning problem, whose goal is to improve the fairness of the commodity flows obtained by using the MMCF based partitioning scheme. This problem can be stated as follows.

{\bf Fair Partitioning Problem:}\label{sec:fpp} {\em Given a solution for the VPN-CS problem with individual commodity flows $f_{mc}(k), \forall k \in K$, the fair partitioning problem is intended to rearrange the path flows for each commodity so as to minimize  $|f_{mc}(i)/\alpha(i) - f_{mc}(j)/\alpha(j)|, \forall i,j \in K$.}

The goal of the fair partitioning problem is to minimize the difference between the ratio of flow achieved from the MMCF solution and the maximum flow between any pair of commodities. In order to achieve this, we propose modified MMCF formulations called the bounded MMCF formulations suitable for offline implementation and a flow balancing algorithm for online implementation in dynamic environments.

Both the offline and online improvements to be proposed will be based on balancing commodity flows obtained from MMCF based heuristic. We begin by first dividing the set of commodities $K$ into the deficit set $(\Omega_d)$ and the excess set $(\Omega_e)$ based on the fraction of maximum flow $\alpha(s_k ,d_k)$ achieved by the MMCF solution as discussed next. The sets are derived by first normalizing the aggregate MMCF commodity flow $f_{mc}(k)$ with its corresponding maximum flow $\alpha(k)$. From this, the set $S_K= {f_{mc}(k)/\alpha(k), \forall k\in K}$ is defined. We next define a threshold value $\sigma$ as follows: Let $S_{min}=\min\{S_K\}$ and $S_{max}=max\{S_K\}$. $\sigma$ is set to $(S_{min} + S_{max})/2$. The threshold  $\sigma$  is then used to divide set $K$ into sets $\Omega_e$ and $\Omega_d$ as follows:  $\Omega_d = \{k | k \in K, S_k \le \sigma \}$,  $\Omega_e= \{k | k \in K, S_k > \sigma\}$. Applying this definition of deficit and excess commodity sets, we next discuss the bounded MMCF formulation and the flow balancing algorithm in order to address the fair partitioning problem.
\subsection{Bounded MMCF Formulation}
\label{sec:sec51}

This formulation retains the objective of the MMCF formulation but adds more constraints in the form of lower and upper bounds to the excess and deficit commodity elements with the goal of reducing the imbalance among commodity flows.  Introducing such bounds enables us to achieve a minimum level of fairness for each deficit commodity without allowing the flows of the excess commodities to fall below a certain threshold flow. We represent the upper and lower bound flows of commodity $e \in \Omega_e$ as $\Phi_{e,l}$ and $\Phi_{e,u}$ respectively,  and those for commodity $d\in \Omega_d$ as $\Phi_{d,l}$ and $\Phi_{d,u}$. The modified MMCF formulation applying the new constraints can be stated as follows.

In the formulation, $f_b(k)$ is the aggregate commodity balanced flow resulting from solving the problem. As can be seen from this formulation  the new constraints, i.e. (21) and (22), are the lower and upper bound constraints on the flows of the excess and deficit commodities, whose initialization we discuss next.

\begin{center}{\bf Bounded MMCF formulation for Fair Partitioning Problem:}\end{center}
\vspace{-2px}
{\footnotesize
\begin{alignat}{2}
\nonumber\text{Maximize} \ \  \displaystyle\sum_{k\in K}{f_{b}(k)}& \\
\nonumber\text{Subject to:} \phantom{ \displaystyle\sum_{v_i\in U}\displaystyle\sum_{k\in K}\displaystyle\sum_{(i,j)\in E}x_{i,j}^{v,k}}   \\
\displaystyle\sum_{(i,j)\in E}x_{i,j}^{k} - \displaystyle\sum_{(j,i)\in E}x_{j,i}^{k}  & =  f_{b}(k) && \ \ \forall \ \  k\in K, i=s_k \label{eq:f5_1}\\
\displaystyle\sum_{(i,j)\in E}x_{i,j}^{k} - \displaystyle\sum_{(j,i)\in E}x_{j,i}^{k}   & =  0 && \ \ \forall \ \  k\in K, i \ne s_k, i\ne d_k \label{eq:f5_2}\\
\displaystyle\sum_{(i,j)\in E}x_{i,j}^{k} - \displaystyle\sum_{(j,i)\in E}x_{j,i}^{k}   & =  -f_{b}(k)  && \ \ \forall \ \  k\in K, i=d_k  \label{eq:f5_3}\\
\Phi_{e,l} \le f_b(e) & \le \Phi_{e,u} && \ \ \forall \ \  e\in \Omega_e  \label{eq:f5_4} \\
\Phi_{d,l} \le f_b(d) & \le \Phi_{d,u} && \ \ \forall \ \  e\in \Omega_d  \label{eq:f5_5} \\
\displaystyle\sum_{k\in K}x_{i,j}^{k} &\le t_{i,j} && \ \ \forall \ \  (i,j) \in E \label{eq:f5_6} \\
\nonumber x_{i,j}^{k} \ge 0, \phantom{00} & \beta \ge 0
\end{alignat}
\vspace{-14px}
}

\subsubsection{Lower and Upper Bound Initialization for Bounded MMCF Formulation}
\label{sec:sec511}

The task now is to understand the initialization of these bounds. Care should be taken to define the bounds, so that the solution obtained by the modified bounded MMCF formulation is at least as good as the one obtained by solving the MMCF formulation.  We propose two ways to initialize the lower and upper bounds of the aggregate flow of the commodities in the sets $\Omega_{e}$ and $\Omega_{u}$.

In the first approach, which we identify as the MB-1 formulation, we assume that the maximum flows and MMCF flows for the commodities in set $K$ with respect to the graph $G(V,E)$ are known. In this case, the lower bound $\Phi_{e,l}$ for the excess commodity element $e \in \Omega_e$  is set so that its aggregate flow does not become less than $\sigma\times\alpha(e)$, ($\alpha(k)$ is the maximum flow of commodity pair $(s_k,d_k)$), and the upper bound $\Phi_{e,u}$ does not exceed the flow obtained from the MMCF solution, which is $f_{mc}(e)$. In case of the deficit commodity elements $d\in\Omega_d$, the lower and upper bound settings are reversed. In this case, the lower bound $\Phi_{d,l}$ is set so that the flow achieved is at least $f_{mc}(d)$, and the upper bound $\Phi_{d,u}$  is set so that it does not exceed the desired threshold  $\sigma\times\alpha(d)$.

The second proposal to initialize the bounds
based on the observation that the optimal commodity flow solution of a MConF problem, which ensures fairness through its throughput factor $\beta$, is also a feasible MMCF solution. For the MB-2 formulation, we must first determine the throughput factor $\beta$ by solving the MConF problem with the demands set to maximum flow of the commodity $k, \alpha(k)$. Once the MConF flows are obtained, the lower bound $\Phi_{e,l}$ of excess commodity $e\in\Omega_e$ is set to $\beta\times\alpha(e)$ and upper bound $\Phi_{e,u}$ to $ f_{mc}(e)$. For the commodity elements in $d\in\Omega_d$, we set lower bound $\Phi_{d,l}$ to  $\beta\times\alpha(d)$, and the upper bound $\Phi_{d,u}$ is set to $\alpha(d)$. We study the performance of the bounded MMCF formulations in the performance analysis section in Section~\ref{sec:sec6}.

The above changes to the MMCF formulation are expected to improve fairness among the commodity flows, but as stated earlier, the formulations would have to be solved in an offline manner using an LP tool, which would make the approach too complex to be realized in a dynamic environments. So, we next discuss a heuristic for online implementation.

\subsubsection{Flow Balancing Heuristic}
\label{sec:sec512}

\begin{figure}[tph]
\begin{center}
\includegraphics[keepaspectratio, scale=0.6]{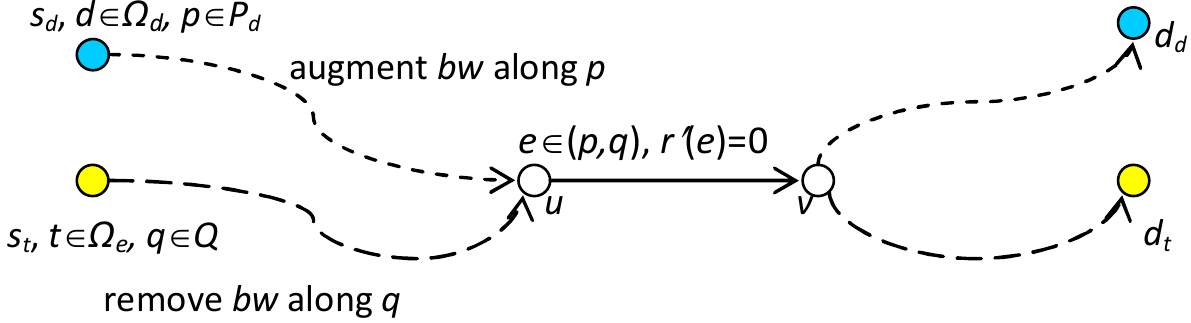} \caption{Logical view of the flow balancing process} \label{fig:3}
\end{center}
\end{figure}

This algorithm builds on the approach discussed in the previous section where the set of commodities $K$ is first divided into an excess set $\Omega_e$ and a deficit set $\Omega_d$. The two sets are determined similarly to the approach discussed earlier using the threshold factor $\sigma$. The algorithm builds on the idea of transferring flows from the elements in excess commodity set $\Omega_e$ to the elements in deficit commodity set $\Omega_d$. The algorithm has as inputs the MMCF flow $f_{mc}(k)$ and the corresponding path set $P_k$  for commodity $k\in K$, and it outputs the modified balanced flows $f_b(k)$ for commodities in the  set $K$. The flow balancing algorithm is explained next with reference to Fig.~\ref{fig:3}
and the pseudo code in Algorithm 3.

In Lines 4--5
, the algorithm begins by identifying the sets $\Omega_e$ and $\Omega_d$ and initializing $f_b(k)$ to the previously computed $f_{mc}(k)$. Flow balancing begins by iterating through each deficit commodity item $d \in \Omega_d$ in Line 6.
In Line 7
, for each commodity d, we iterate over all the paths $p\in P_d$ corresponding to the commodity flow $f_b(d)$. We choose a path $p = \{S_d, e_1, e_2,\ldots e, \ldots D_d\}$ for further steps in the iteration only if it has exactly one edge $e$ such that the left over edge capacity after executing MMCF based partitioning algorithm, which we denote as $r^\prime(e)$, satisfies $r^\prime(e)=0$, and $r^\prime(e_i)\ge\tau$ for all other edges $e_i \in p$ and $e_i\ne e$, for a predefined left over edge capacity threshold $\tau$. We impose this condition for three reasons: First, the edge with the zero left over capacity increases the probability of finding at least one or more commodities from the $\Omega_e$ set contributing to the total edge flow, and this could be used to potentially transfer flow to the deficit commodity; second, this approach enables us to control the complexity of the algorithm by limiting the choice of edges that can be chosen on a path $p\in P_d$, and thereby, simplifying the choice of the excess commodity from which the flow can be transferred; third, $\tau$ also acts like a tunable parameter that could also be used to control the complexity of the algorithm by eliminating the paths that may not make a considerable change to the flow of the deficit commodity after an augmentation iteration. If an FPTAS ( Fully Polynomial Time Approximation Scheme) algorithm was used to solve the MMCF problem, and as the optimality of the solution is a function of the approximation factor, the left over capacity on all the edges of the path p may be positive. In this case, we chose the path $p$ and an edge e such that $e \in \{x: \min\{r^\prime(x), x\in p\}\}$ and with left over capacity $r^\prime(e_i) > \tau, \forall e_i \in p$ and $e_j \ne e$.

\begin{algorithm}[tph]
{\small
\begin{algorithmic}[1]
\REQUIRE{ $G = ( V, E)$, $K$, $f_{mc}(k)$, $\alpha(l)$,$P_k\ \ \forall\ k \in K$, $\sigma$, $\tau$}
\ENSURE{$f_b(k) \forall \ k \in K$ and modified $x_{i,j}^k$, $\forall \ (i,j) \in E$}
\FORALL{commidity $k\in K$}
	\STATE{set $f_b(k) = f_{mc}(k)$}
\ENDFOR
\STATE{Partition the set of commodities into two sets, $\Omega_e$ and $\Omega_d$, using threshold factor $\sigma$}
\STATE{Identify path set $P_d$, $\forall d \in \Omega_d$ and path set $P_e$, $\forall e \in \Omega_e$. Initialize for $\forall k \in K$, $f_b(k) = f_{mc}(k)$}
\STATE{Iterate line $7$ -- $16$ over each deficit commodity $d$, $\forall d \in \Omega_d$}
\STATE{Iterate line $8$ -- $16$ over each path $p \in P_d$ with only one saturated link $e \in P_d$ in the path, where $e = \{x: min\{r^\prime(x), x\in p\}\}$ and with non-zero left over capacities in the remaining links i.e. $r^\prime(e_i)>\tau,\ \forall e_i\in p$ and $e_i \neq e$}
\STATE{Find the subset $T \subset \Omega_e$ of excess commodities having non-zero flow on the link $e \in P_d$}
\STATE{From the set $T$, we determine the commodity $t \in T$, which can contribute maximum additional flow. This is done by choosing the commodity t that maximizes the difference $\frac{f_b(t)}{\alpha(t) - \sigma}$}
\STATE{Once the excess commodity $t$ is chosen, identify a path set $Q =\{x: x\in P_t, e \in x\}$. From $Q$, choose $q\in Q$ that maximizes the flow of commodity $t$ on path $q$}
\STATE{The amount of flow that can be transferred from excess commodity t to the deficit commodity d is determined by: $bw = min \{\lambda_{t,q} ,\zeta_{d,p}, f_d , f_t \}$}
\COMMENT{Note: here, $\lambda_{t,q}$ is the flow of commodity $t$ along path $q$; $\zeta_{d,p}$ is the minimum of left over capacity of path p corresponding to deficit commodity $d$; $f_d$ is the flow required for deficit commodity $d$ to achieve its target flow threshold  $\sigma\times\alpha(d)$; $f_t$ is  the maximum flow that can be contributed by  commodity t beyond which the flow of the excess commodity would go below the threshold flow $\sigma\times\alpha(t)$ )}
\STATE{Augment flow equal to $bw$ along the deficit commodity path $p$. The equivalent flow is also deduced from excess commodity path $q$. Also, update $f_b(d)= f_b(d) + bw$ and  $f_b(t)= f_b(t) - bw$. If $(f_b(t) - \sigma_{th} \alpha(t)) \le 0$, update set $\Omega_e = \Omega_e - \{t\}$}
\IF{ $\Omega_e=0$}
	\STATE{Go to end}
\ENDIF
\STATE{Check to see if deficit commodity has attained the threshold flow, i.e.   $(f_b(d) - \sigma_{th} \alpha(d)) \ge0$ . If so, the heuristic proceeds to consider the next deficit commodity (line $6$), else repeats the process for next path in the set $P_d$ (line $7$)}
\caption{Flow balancing heuristic for fair partitioning problem}
\label{alg:fbhfpp}
\end{algorithmic}
}
\end{algorithm}

Once a path $p$ is determined, we determine in Line 8
the set $T \subset \Omega_e$ of excess commodities that can potentially contribute flows to the deficit commodity $d$. In Line 9
, we select an excess commodity $t\in T$ that can contribute the maximum units of flow. We choose $t \in T$, which maximizes $(f_b(t)/\alpha(t)- \alpha)$, i.e $t \in \{x: \max\{ f_b(x)/\alpha(x) - \alpha \} \wedge x\in T \}$. In Line 10
, corresponding to the commodity $t$, we identify the subset $Q$ of commodity path set $P_t$ that includes edge $e$. From the set $Q$, we choose the path $q\in Q$ that maximizes the flow of commodity $t$ on path $q$; we represent this flow as $\lambda_{t,q}$. In Line 11
, to determine the flow that can be transferred from the excess commodity t to the deficit commodity $d$, we compute the following:

\begin{itemize}
\item $\lambda_{t,q} =$ Minimum of the flows of the excess commodity $t$ on the edges of path $q$.
\item $\zeta_{d,p} =$ Minimum of the left over capacities of the links on path $p$ of the deficit commodity $d$ without considering the left over capacity of the link $e$.
\item $f_d= (\sigma \times\alpha(d) - f_b(d))$ is the flow required for deficit commodity $d$ in order to achieve its target flow threshold,  which is $\sigma\times\alpha(d)$.
\item $f_t = (f_b(t) - \sigma\times\alpha(t))$ is  the maximum flow that can be contributed by  commodity t beyond which the flow of the excess commodity would go below the threshold flow $\sigma\times\alpha(t)$.
\end{itemize}

We choose the minimum of the above values in order to determine $bw$, i.e    $bw=min\{\lambda_{t,q}, \zeta_{d,p}, (\sigma\times\alpha(d) - f_b(d)), (f_b(t) - \sigma\times\alpha(t))\}$. In Lines 12--15
, $bw$ is deducted from edge flows corresponding to the excess commodity t on path q and augmented along the edges of path p of the deficit commodity $d$. Corresponding augmentation and subtraction are also made to respective aggregate commodity flows $f_b(d), f_b(t)$, and set $\Omega_e$  is updated if necessary. If $\Omega_e$ is empty, we terminate, else this process continues over all the deficit commodities until either each of the deficit commodities $d\in\Omega_d$ satisfies $(f_b(d)/\alpha(d))\ge \sigma$, or there are no more paths $p$ for augmentation in the set $P_d$  in the iteration for commodity $d$, upon which we move on to the next deficit commodity in the set $\Omega_d$.

The complexity of the balancing heuristic can be controlled by limiting the number of paths considered for each commodity in $\Omega_d$ and $\Omega_e$. The worst case complexity of the flow balancing heuristic can be derived assuming a highly unfair flow distribution by MMCF, where the size of $|\Omega_d|$ is $O(|K|)$. Let $|P_d|$ and $|P_e|$ be the maximum number of paths to be considered for each deficit and excess commodity as part of the balancing process. The worst case complexity of the flow balancing heuristic with this assumption is $O(|K|\times|P_d|\times|P_e|)$.

The flow balancing heuristic can be incorporated as part of the MMCF based partitioning heuristic in order to improve the fairness of the MMCF commodity flow by modifying Line 5
(Algorithm 2)
as follows:

Line 5
: For the commodity set $K$, solve the maximum multicommodity flow problem.  The solution $x_{i,j}^k$   represents the resource associated with commodity $k$ over edge $(i,j)\in E$.

Line 5a
: If the MMCF fair partitioning scheme is enabled, determine set  $S_K= \{f_{mc}(k)/ \alpha(k)\}, \forall k \in K$ , Initialize $\sigma$;

Line 5b
:  Apply MB-1, MB-2, or flow balancing heuristic in order to obtain new $x_{i,j}^k$'s and balanced MMCF flows $f_b(k)$ which is the new  $f_{mc}(k)$;

The complexity of the MMCF based partitioning heuristic for the online implementation case with the above changes could be dominated by  the complexity of solving  MMCF formulation to optimality or   $\epsilon$-approximation algorithm or  $O(|K|)$ max flows computation and flow balancing algorithm. Considering this, the overall pseudo-polynomial complexity using FPTAS for MMCF proposed in~\cite{lisa1999}, with the previously discussed flow balancing algorithm, is $O(\epsilon^2|E|((|E|+|V|\times \log|E|)\log(|V|))+(|K|(|V|^3+(|P_d|\times|P_e|) ))$.
\section{Simulation Study}
\label{sec:sec6}

\begin{figure}[tph]
\begin{center}
\includegraphics[height=35mm]{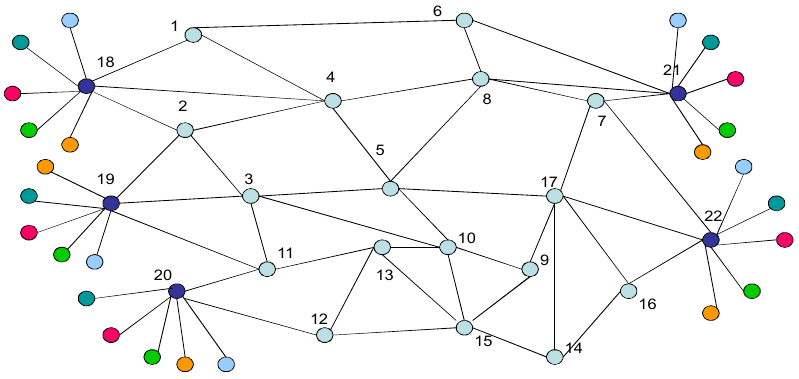} \caption{Simulation topology} \label{fig:4}
\vspace{-2.05em}
\end{center}
\end{figure}

The topology used for studying the different scenarios is a $22$-node random topology based on well-known Waxman's random graph~\cite{waxman1988} model with an average node degree of $4$, $\alpha=0.150,\ \beta =2.2$; this topology is shown in Fig.~\ref{fig:4}.
In order to validate our results over other standard topologies, we also studied the performance with respect to two other European networks referred from~\cite{kwok}. Considering the correlation in the results for the three topologies, for the discrete event simulation case we limit our results to those corresponding to the random graph. All the results discussed have been obtained from running the simulation for $30$ independent replications to achieve $95\%$ confidence interval for an {\em absolute error}~\cite{averill2006} of less than $1\%$. The number of independent replications is deduced by observing the sample variance over several independent runs, and applying the approximation given in \cite{averill2006} (page. 512). The simulation study has been carried out with respect to three objectives.

{\footnotesize
\begin{table}[tphb]
  \centering
    \caption{Simulation parameters}\label{tbl:2}
    \footnotesize
    \begin{tabular}{|p{0.44\columnwidth}|p{0.12\columnwidth}|p{0.12\columnwidth}|p{0.12\columnwidth}|}
    \hline
    {Simulation Parameters} & Objs. 1-2 & Obj. 3(a) & Obj. 3(b) \\\hline
    {VPN Mean Call Inter-arrival time (s)} & 100   & 100   & 100 \\\hline
    {VPN Mean Call Holding Time ($\mathcal{H}$) (s)} & [10,1000] & [10,1000] & [50,1000] \\\hline
    {Core Topology Link state Update Interval (s)} & 5     & 5     & 5 \\\hline
    {CS Abstract Topology Update Interval (s)} & 10    & 10    & 10 \\\hline
    {VPN Abstract Topology Type (VPN (A-E))} & SSA   & SSA   & SSA \\\hline
    {Abstract Topology Refresh Interval (VPNA) (s)} & 100   & 100   & 100 \\\hline
    {Abstract Topology Refresh Interval (VPNB) (s)} & 100   & 100   & 100 \\\hline
    {Abstract Topology Refresh Interval (VPNC) (s)} & 100   & 100   & 100 \\\hline
    {Abstract Topology Refresh Interval (VPND) (s)} & 100   & 100   & 100 \\\hline
    {Abstract Topology Refresh Interval (VPNE) (s)} & 100   & 100   & 100 \\\hline
    {Approximation factor (for MConF/MMCF FPTAS)} & 0.7   & 0.7   & 0.7 \\\hline
    \end{tabular}%
  \label{tab:addlabel}%
\end{table}}%

\begin{figure*}[h]
\centering
\subfloat[]{\includegraphics[width=.33\textwidth, height=45mm]{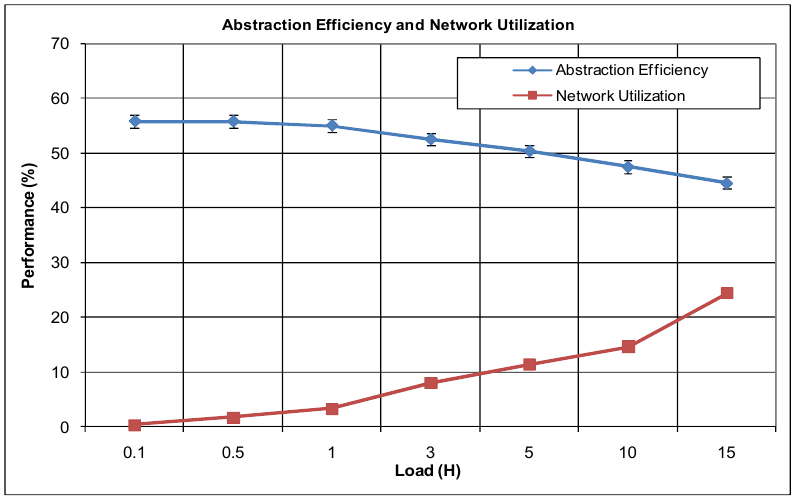}\label{fig:5}}
\subfloat[]{\includegraphics[width=.33\textwidth, height=45mm]{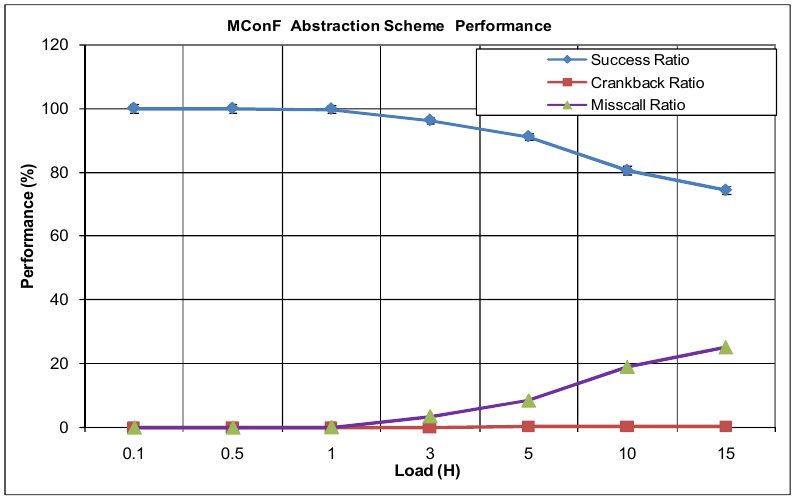}\label{fig:6}}
\subfloat[]{\includegraphics[width=.33\textwidth, height=45mm]{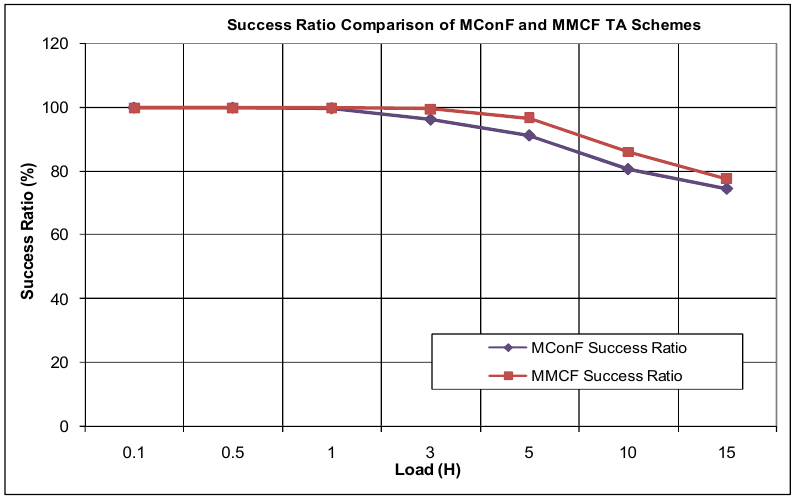}\label{fig:7}}
\caption[]{\subref{fig:5} VSP abstraction efficiency and core network utilization with varying mean holding time ($\mathcal{H}$); \subref{fig:6} VPN call performance with varying $\mathcal{H}$ for MConF based abstraction scheme; \subref{fig:7} VPN call performance comparison of success ratio MConF and MMCF based abstraction schemes with varying $\mathcal{H}$.}
\end{figure*}

{\bf Objective 1(a-b)}:  First we study the performance of the MConF based resource partitioning algorithm. Then, we study and compare the performance of the MMCF based resource partitioning algorithm with respect to the MConF based scheme.

{\bf Objective 2(a-b)}: As part of this objective, in the first part, we study the performance with respect to the improvement achieved by the bounded MMCF formulations (MB-1 and MB-2) in terms of fairness and compare their performance to the performance of the MMCF formulation without the fairness constraints. The analysis is done in an offline manner using an LP tool~\cite{lindo} over well known service provider networks referred from~\cite{kwok}. In the second part, we present a simulation analysis of the performance of the flow balancing algorithm proposed to improve the fairness of MMCF based partitioning scheme.

{\bf Objective 3}: The goal of this objective is to compare the performance of the centralized abstraction schemes with the decentralized schemes proposed in~\cite{ravi2013}. Here, we also study how oversubscription of the residual link capacity can be used to overcome the conservative nature of centralized TA generation schemes to a certain extent. We also study the effect of oversubscription on VPN call performance and network utilization.
In the simulation study, all final abstractions from the partition subgraphs are derived using the maximum capacity scheme of~\cite{ravi2013}. We choose maximum capacity scheme as it performed the best in terms of success ratio of the three schemes proposed in~\cite{ravi2013}.

Figures~\ref{fig:5}--\ref{fig:6}
show the performance of MConF based subgraph partitioning scheme for generating VPN TA by varying the mean call holding time ($\mathcal{H}$). With respect to abstraction efficiency (Fig.~\ref{fig:5}),
the MConF based algorithm achieved an abstraction efficiency of $55\%$ at lower load conditions, which decreases by $10\%$ with an increasing load on the network. This is because increasing the mean call holding time of VPN bandwidth requests causes the network utilization to increase, as shown in Fig.~\ref{fig:6}. 
 Hence, the residual capacity for future abstraction reduces. This results in an increasing variance of residual capacity among the links resulting in lesser aggregate commodity flows, and causing the resource partitioning scheme to result in lower abstraction efficiency.

\begin{figure*}
\centering
\subfloat[]{\includegraphics[width=.67\columnwidth, height=45mm]{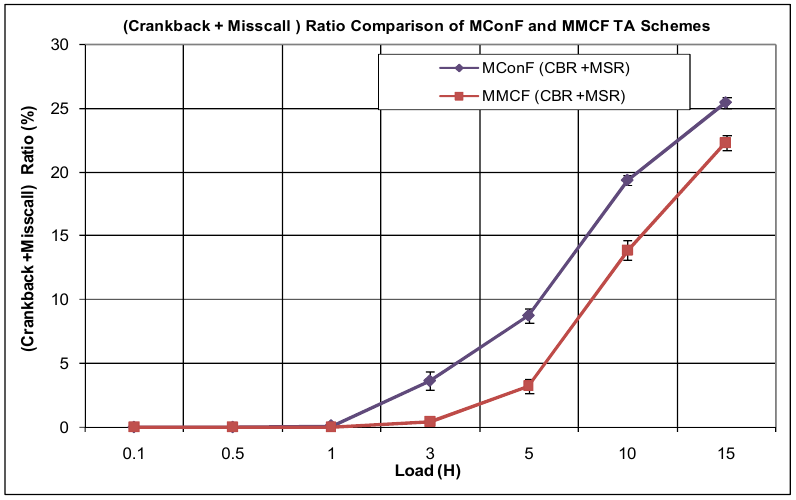}\label{fig:8}}
\subfloat[]{\includegraphics[width=.67\columnwidth, height=45mm]{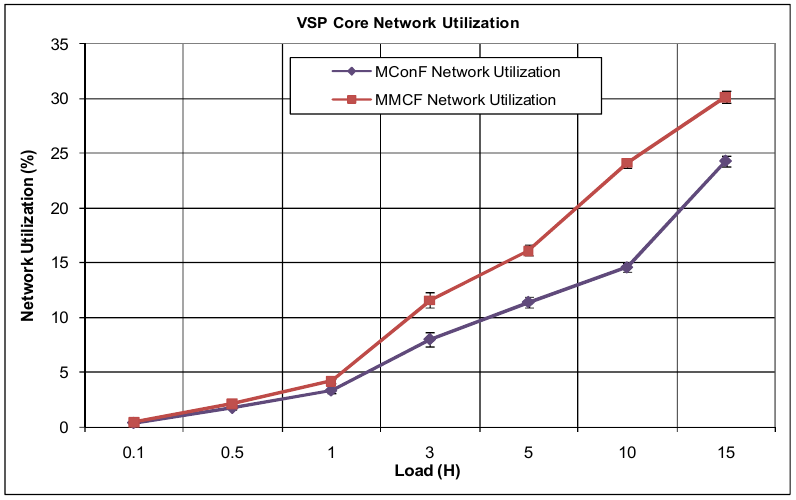}\label{fig:9}}
\subfloat[]{\includegraphics[width=.67\columnwidth, height=45mm]{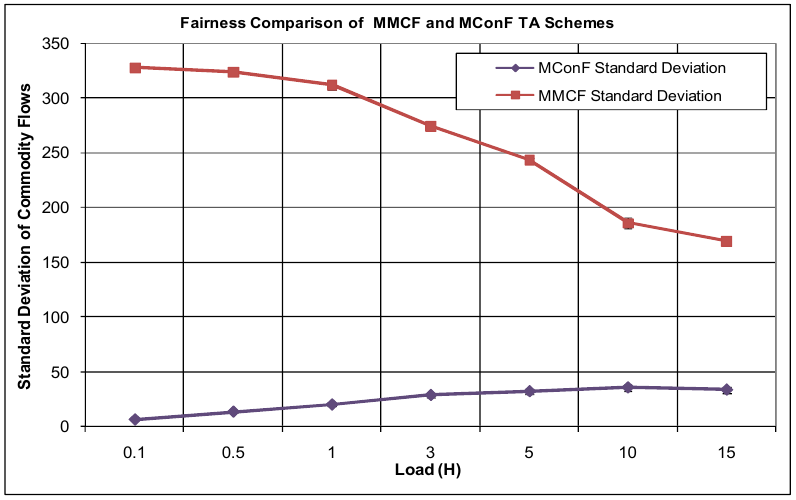}\label{fig:10}}\\
\subfloat[]{\includegraphics[width=.67\columnwidth, height=45mm]{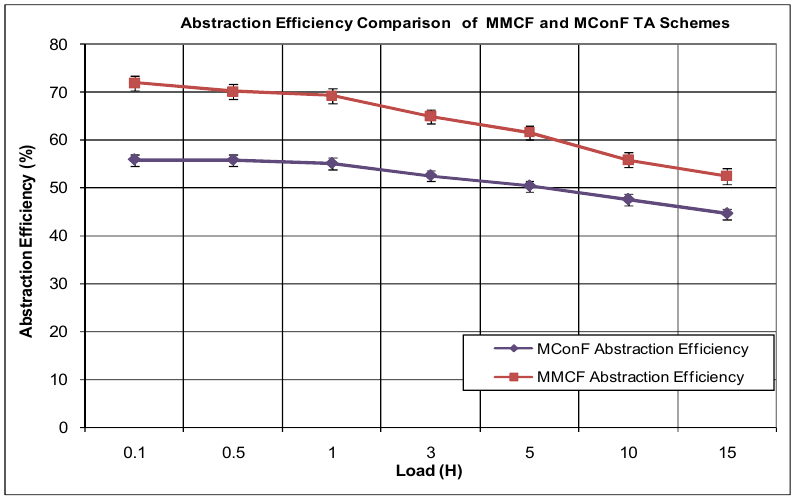}\label{fig:11}}
\subfloat[]{\includegraphics[width=.67\columnwidth, height=45mm]{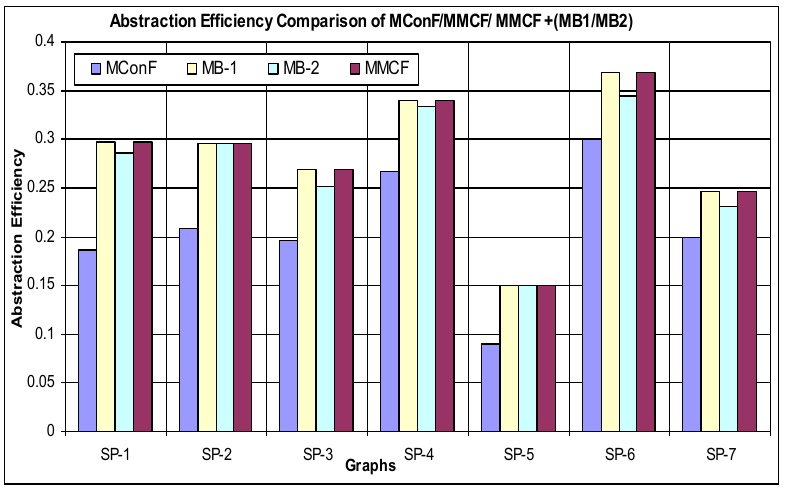}\label{fig:12}}
\subfloat[]{\includegraphics[width=.67\columnwidth, height=45mm]{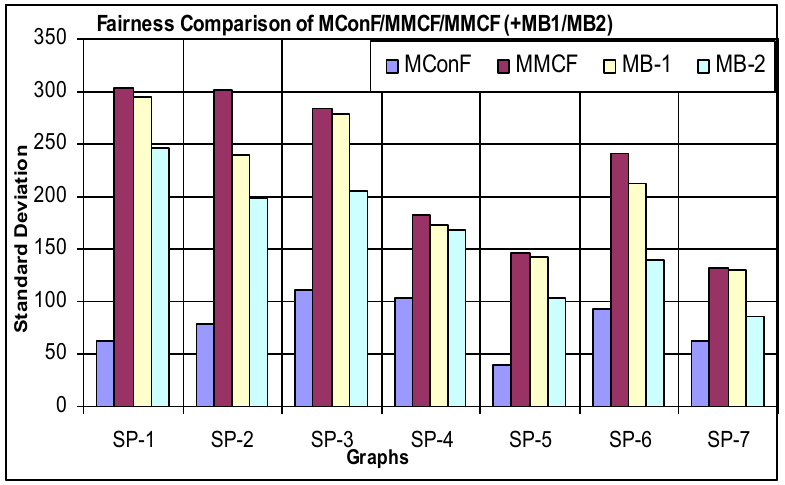}\label{fig:13}}
\caption[]{\subref{fig:8} VPN call performance (crankback + misscall) ratio comparison of MConF and MMCF based abstraction schemes with varying mean holding time ($\mathcal{H}$); \subref{fig:9} VSP network utilization comparison of MConF and MMCF based abstraction schemes with varying $\mathcal{H}$; \subref{fig:10} Comparison of fairness of MConF and MMCF based abstraction schemes with varying $\mathcal{H}$; \subref{fig:11} Comparison of abstraction efficiency of MConF and MMCF based abstraction schemes with varying $\mathcal{H}$; \subref{fig:12} Offline comparison of abstraction efficiency of MConF, MMF, MMCF with MB1/MB2 formulations for different graphs; \subref{fig:13} Offline comparison of fairness of MConF, MMCF, MMCF with MB1/MB2 formulations for different graphs.}
\end{figure*}

{\bf Objective 1(a): Performance of MConF based partitioning Scheme}

With respect to call performance in Fig.~\ref{fig:6},
we observe that at lower load conditions, the success ratio is close to $100\%$, but as the load increases, the success ratio decreases. This is because of increasing misscall ratio. It should be noted that crankback ratio is almost 0 for all the load conditions. The good crankback ratio is achieved because of the subgraph partitioning method used for TA generation, but this is at the cost of deteriorating misscall ratio performance. The main reason for misscall ratio to increase with increasing load is due to the reducing abstraction efficiency as noted earlier. With lesser abstraction efficiency, the virtual links are associated with lesser virtual link capacities. This makes the VPNs terminate increasing percentage of calls locally, thereby increasing the misscall ratio.

In conclusion, we observe that applying MConF based TA generation approach achieves very good success ratio at low load conditions, but at higher loads the success ratio deteriorates because of  poor misscall ratio performance. Also, applying centralized mode of TA generation results in very good crankback ratio.

{\bf Objective 1(b): Evaluation of  the MMCF based partitioning scheme and comparison with the MConF based scheme}

Figures~\ref{fig:7}--\ref{fig:8}
compare the call performance of MConF and MMCF based heuristics under varying loads. From Fig.~\ref{fig:7},
which compares the success ratio, we observe that the partitioning approach using MMCF performs better than MConF by about $5\%$ particularly at higher load conditions. This is because of the gain achieved by MMCF in terms of abstraction efficiency. A similar result was also made in the case of crankback and misscall ratio metrics as shown in Fig.~\ref{fig:8}. 
We can see that MMCF based TA scheme performs $5\%$ better than the MConF based TA scheme in terms of this aggregate metric. Figure~\ref{fig:9} 
compares the network utilization achieved by the MConF and MMCF approaches; the better call performance and abstraction efficiency of the MMCF based approach results in better network utilization in the range of $5\%$ for various load conditions. As expected, a significant difference in the two approaches was noted with respect to fairness of the commodity flows. Though MMCF was observed to achieve high abstraction efficiency, it fared poorly in terms of fairness among the commodity flows. Figure~\ref{fig:10} 
compares the average standard deviation of the commodity flows achieved by the MMCF and MConF based partitioning algorithms. We see that at lower network utilization conditions, the MConF performs fifty times better than MMCF. The decreasing difference of the fairness performance of the two schemes is due to decreasing difference of abstraction efficiency between the two schemes with increasing load as observed in Fig.~\ref{fig:11}.

\begin{figure*}[ht]
\centering
\subfloat[]{\includegraphics[width=.33\textwidth, height=45mm]{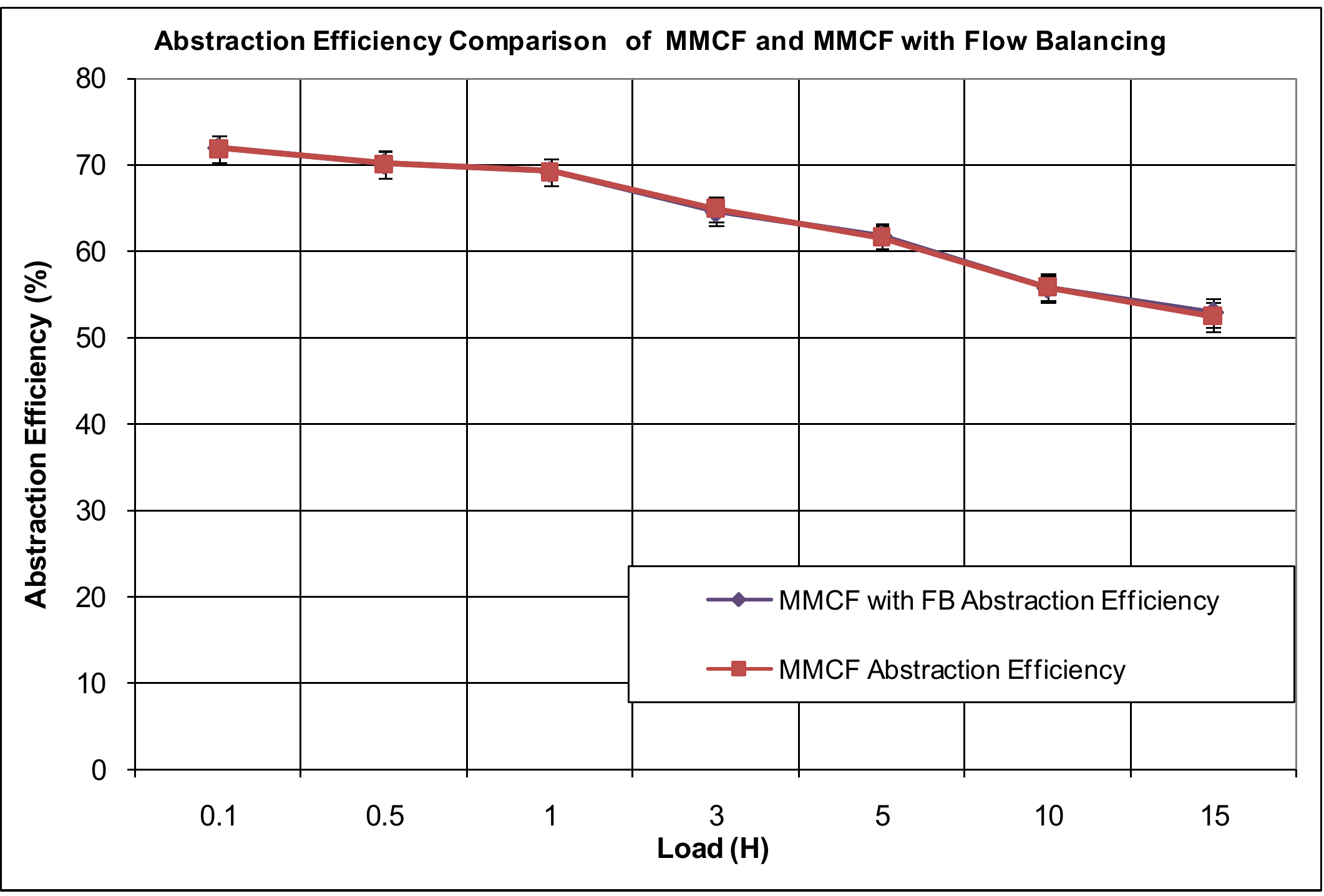} \label{fig:14}}
\subfloat[]{\includegraphics[width=.33\textwidth, height=45mm]{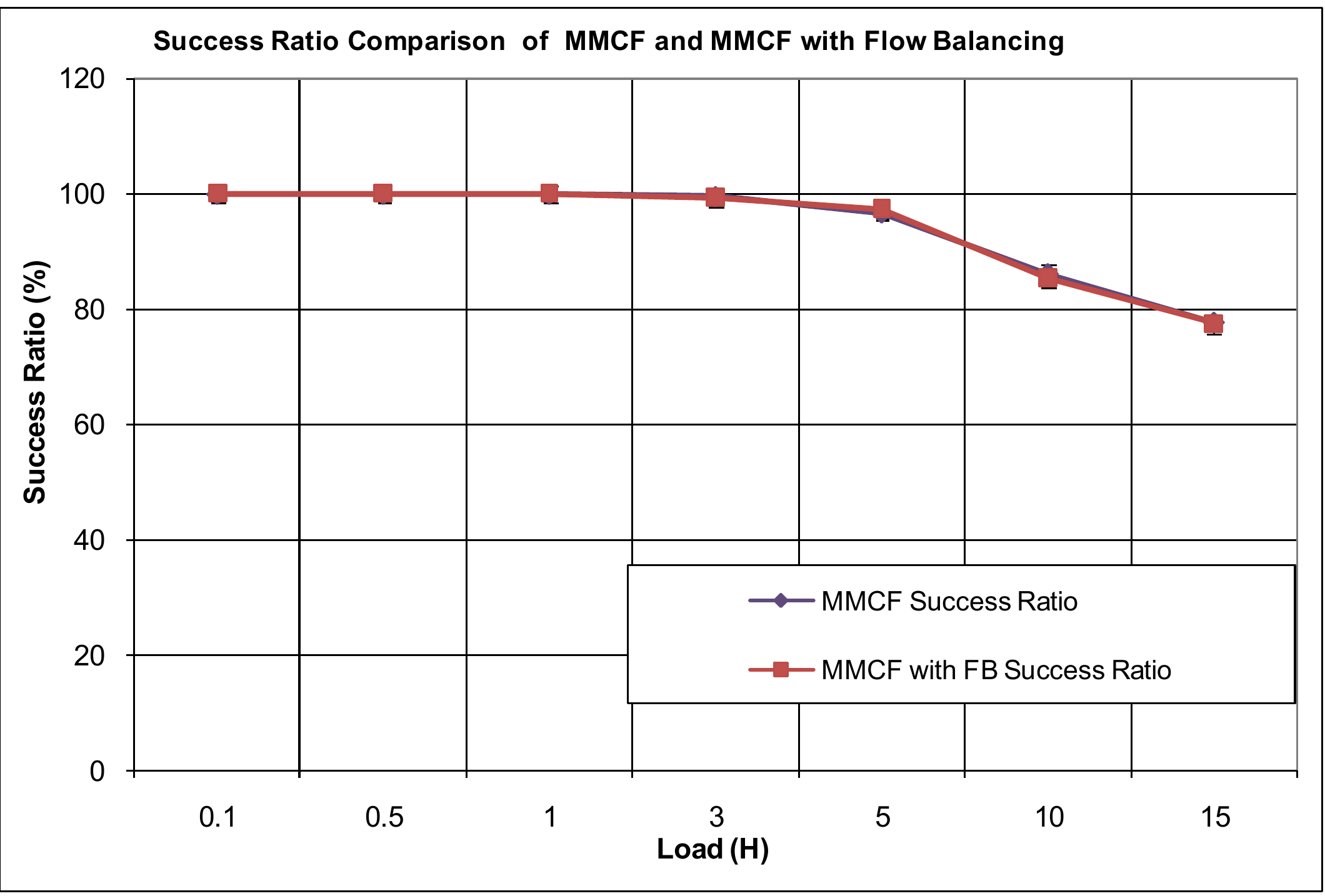}\label{fig:15}}
\subfloat[]{\includegraphics[width=.33\textwidth, height=45mm]{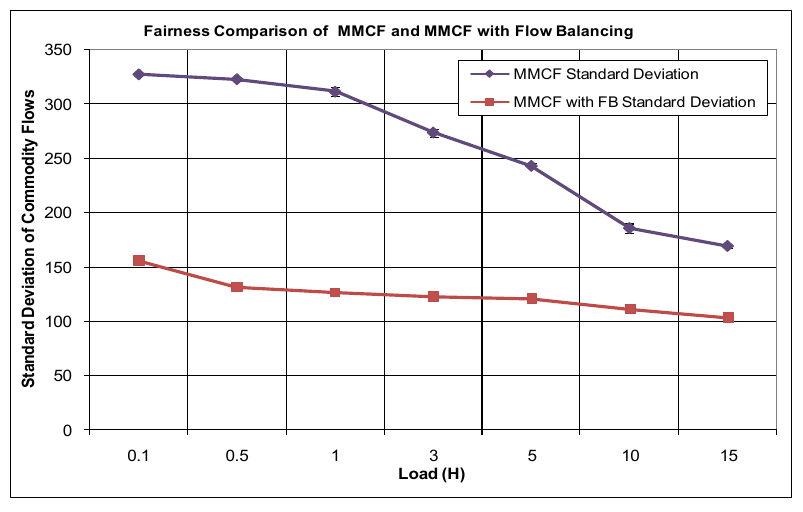} \label{fig:16}}
\caption[]{\small \subref{fig:14} Comparing VSP abstraction efficiency of MMCF with and without Fair Balancing with varying mean holding time ($\mathcal{H}$);  \subref{fig:15} VPN call performance comparing success ratio of MMCF with and without Fair Balancing with varying $\mathcal{H}$; \subref{fig:16} Comparing fairness of MMCF with and without Fair Balancing with varying $\mathcal{H}$.}
\end{figure*}

{\bf Objective 2(a): Offline performance comparison of MMCF formulations and improvements to achieve fairness}

Here, we study and compare the performance of MMCF formulation and its variations to address the fairness issue in an offline manner. Before discussing the results, we wish to note that by the very nature of the objective function, MMCF will always result in the maximum abstraction efficiency and least fairness, and as stated earlier, MConF will result in commodity flows with maximum fairness, while achieving the least abstraction efficiency. The performance of the MMCF variations MB-1 and MB-2 is expected to lie within these extremes. The goal of this objective is to understand which formulation strikes the best balance in achieving abstraction efficiency matching that of the MMCF scheme and fairness matching that of the MConF scheme.  We study the performance of these formulations by running over well known provider networks referred from~\cite{kwok}, and we characterize their performance in terms of two metrics: the aggregate flow achieved by the formulation from which abstraction efficiency is obtained and the fairness of the commodity flows among the commodities. The fairness is characterized in terms of the standard deviation among the commodity flows. In Fig.~\ref{fig:12},
we see that MMCF achieves the best abstraction efficiency but also suffers in terms of fairness as shown in Fig.~\ref{fig:13},
which results in maximum standard deviation among the commodity flows. On the other hand, the MConF resource partitioning scheme achieves the least abstraction efficiency, while achieving the least standard deviation and maximum fairness. Comparing the two MMCF variations, variation MB-1, which defines its flow bounds using the maximum flow and MMCF flow, achieves abstraction efficiency close to that of MMCF formulation but makes only a small gain in addressing the fairness issue. The second variation MB-2 that uses the MConF and MMCF outputs to define the bounds results in as much or slightly less aggregate flow when compared to MMCF or MB-1 formulation, while achieving maximum fairness of the three MMCF formulations. Hence we see that, comparing the two bounded MMCF formulations, the MB-2 variation that uses the MConF flows to define the bounds over the commodity flows, performs the best in terms of achieving MMCF's abstraction efficiency, while also improving the fairness among the commodity flows.

\begin{figure*}
\centering
\subfloat[]{\includegraphics[width=.67\columnwidth, height=45mm]{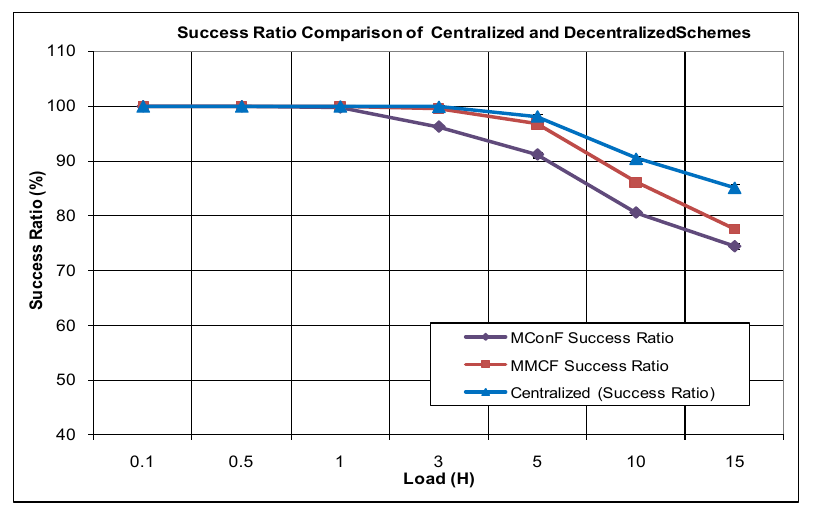} \label{fig:17}}
\subfloat[]{\includegraphics[width=.67\columnwidth, height=45mm]{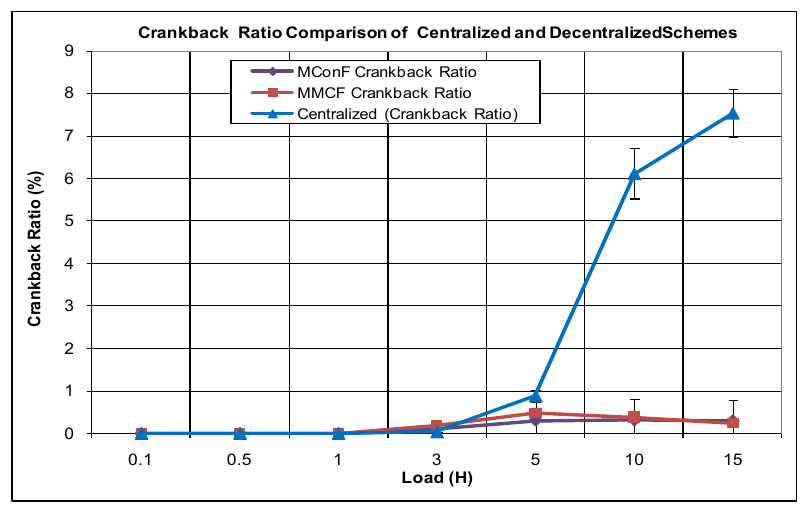} \label{fig:18}}
\subfloat[]{\includegraphics[width=.67\columnwidth, height=45mm]{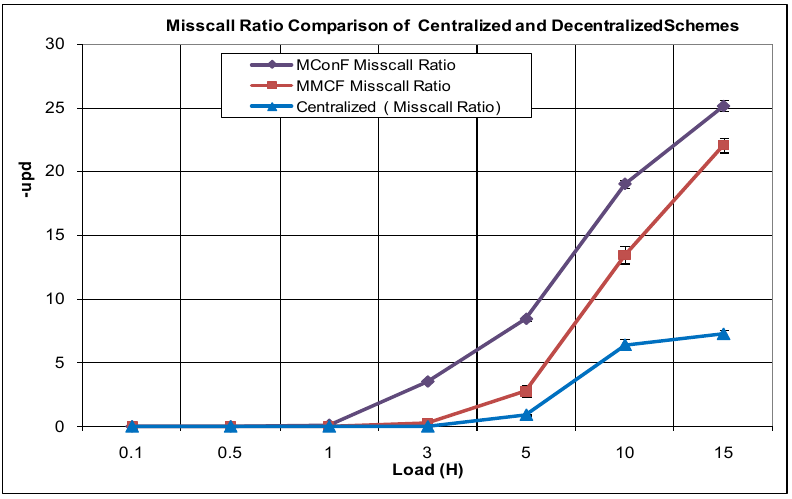} \label{fig:19}}\\
\subfloat[]{\includegraphics[width=.67\columnwidth, height=45mm]{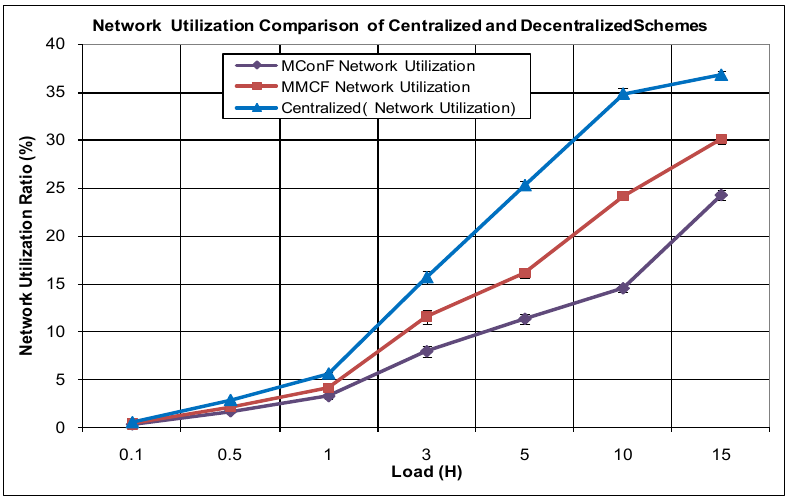} \label{fig:20}}
\subfloat[]{\includegraphics[width=.67\columnwidth, height=45mm]{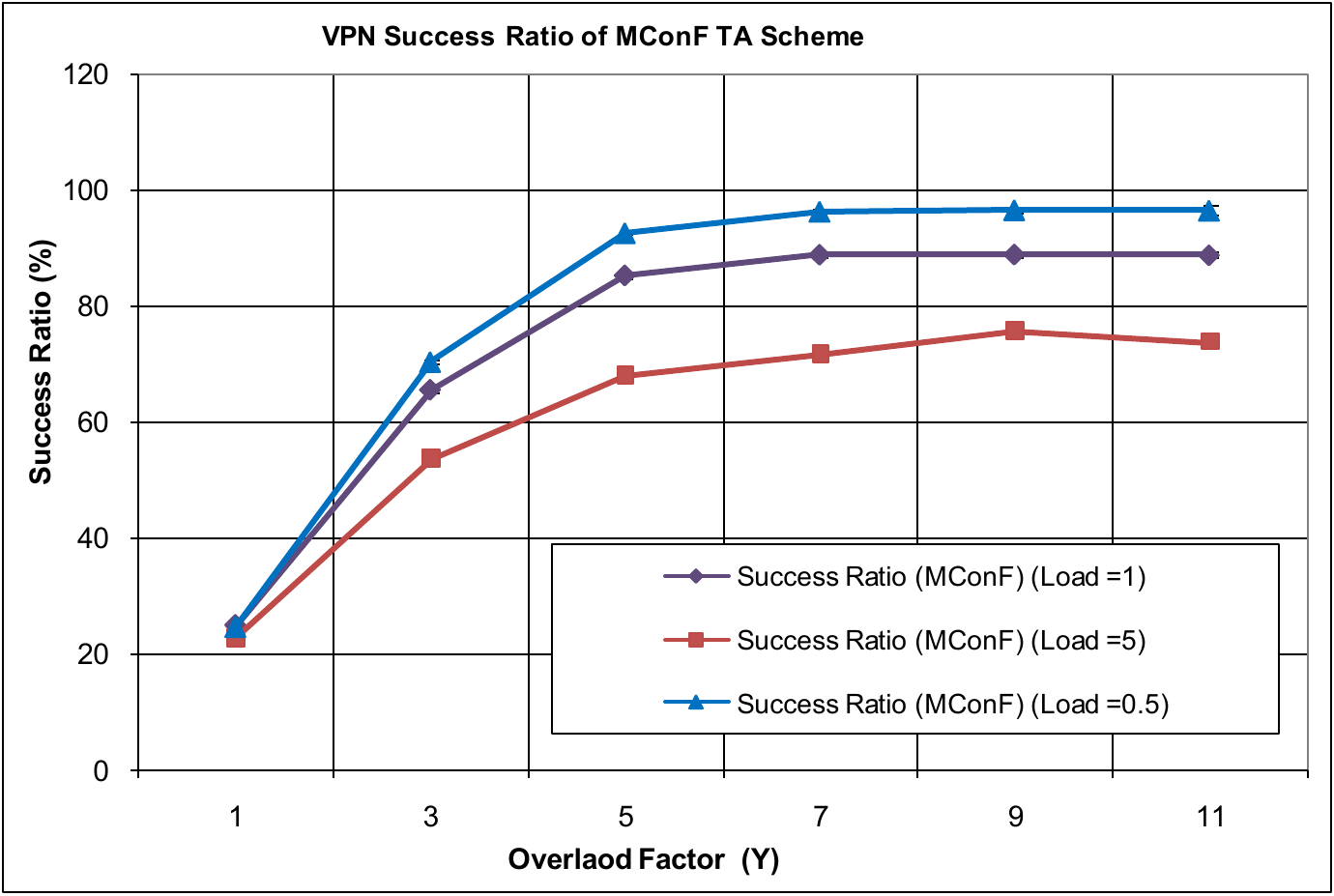} \label{fig:21}}
\subfloat[]{\includegraphics[width=.67\columnwidth, height=45mm]{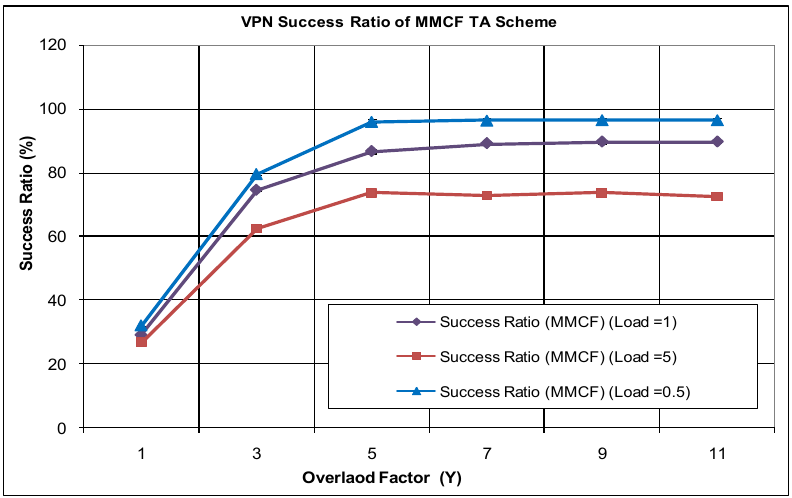} \label{fig:22}}
\caption[]{\small \subref{fig:17} VPN call performance comparing success ratio of decentralized , MConF and MMCF schemes with varying mean holding time ($\mathcal{H}$); \subref{fig:18} VPN call performance comparing crankback ratio of decentralized , MConF and MMCF schemes with varying $\mathcal{H}$; \subref{fig:19} VPN misscall ratio comparison of decentralized , MConF and MMCF schemes with varying $\mathcal{H}$; \subref{fig:20} VSP network utilization comparison of decentralized , MConF and MMCF schemes with varying $\mathcal{H}$; \subref{fig:21} VPN call performance comparing success ratio of  MConF based abstraction scheme with varying over-subscription factor (Y) with 95\% C.I; \subref{fig:22} VPN call performance comparing success ratio of  MMCF based abstraction scheme with varying oversubscription factor (Y).}
\end{figure*}

{\bf Objective 2(b): Comparing performance of MMCF based partitioning scheme and MMCF augmented with flow balancing heuristic}

Figures~\ref{fig:14}--\ref{fig:15}
compare the performance of abstraction efficiency and success ratio of MMCF with and without the application of the flow balancing heuristic. With respect to both these metrics, the performance of the two scenarios was almost the same. No difference in performance was also noted in terms of other call performance metrics too. This is expected since the fair partitioning algorithm only improves the fairness of the commodity flows obtained from MMCF based heuristic, which  is done by rebalancing the flows without affecting the abstraction efficiency.

As expected, an important difference was noted with respect to fairness, as seen in Fig.~\ref{fig:16}.
With flow balancing enabled over MMCF commodity flows, we would expect the variance of the commodity throughput factor to reduce. The improvement in terms of standard deviation when flow balancing is applied was noted to be about $25\%$ better; this demonstrates the usefulness of the flow balancing heuristic. From Fig.~\ref{fig:16}, 
we observe that with increasing load the difference between the variance of the commodity flows of MMCF and MMCF with flow balancing decreases. This is because, the increasing network load results in proportional increase in network utilization, hence decrease in abstraction efficiency. This results in lesser aggregate MMCF flows, thereby decreasing the variance between aggregate flows of the commodities. This causes flow balancing scheme to achieve lesser improvement in terms of fairness with increasing load.

{\bf Objective 3(a): Comparing Performance of centralized abstraction and decentralized abstraction schemes}

In this objective, we compare the difference in performance of the two  modes of TA generation in terms of VPN call performance and network utilization. From the three schemes proposed in~\cite{ravi2013}, we choose the maximum capacity scheme for the analysis. In the following, decentralized scheme refers to the decentralized scheme of~\cite{ravi2013} based on the maximum capacity abstraction algorithm. Similarly, MConF and MMCF schemes, also called  centralized schemes refer to those that construct partition subgraphs using the MCoN and MMCF formulations and then apply the maximum capacity abstraction algorithm.

Figures~\ref{fig:17}--\ref{fig:19}
compare the call performance and network utilization of the decentralized scheme, MConF, and MMCF based abstraction schemes. As regards  the VPN success ratio, we note in Fig.~\ref{fig:17}
that the success ratio of the decentralized scheme scheme is better than that of MConF scheme by $10\%$ and by $5\%$ with respect to the MMCF  scheme.  With respect to the crankback ratio we note in Fig.~\ref{fig:18} 
that the decentralized  scheme's performance degrades by $7\%$ at maximum load while MConF and MMCF  schemes achieve crankback ratio of less than $1\%$, which was also noted earlier in objectives 1(a-b).  The significant gain in crankback ratio is achieved at the cost of high misscall ratio of $25\%$ and $22\%$ for MConF and MMCF  schemes compared to $7\%$ for the decentralized scheme. This performance difference is shown in Fig.~\ref{fig:19}. 
The performance difference in success ratio also translates into good network utilization of the decentralized scheme, which  performs $10\%$ and $5\%$ better than MConF and MMCF  schemes. This can be seen in Fig.~\ref{fig:20}. 
This performance difference between the decentralized and  MConF and MMCF  schemes  suggests that minimizing the oversubscription alone, which was the main motivation for developing the MConF and MMCF  schemes  , does not translate into good VPN call performance. This is particularly obvious at higher loads when the contention of resources among the VPNs is high. In this case, the aggressive nature of the decentralized scheme is noted to improve statistical multiplexing of core resources among the VPN bandwidth requests, which contributes to better call performance. The  centralized  schemes, because of their conservative exposure of available capacity (derived from logical partitioning schemes), result in poorer misscall ratio performance because of their property of minimizing oversubscription, which results in limited multiplexing of available capacity.

\begin{figure*}
\centering
\subfloat[]{\includegraphics[width=.67\columnwidth, height=45mm]{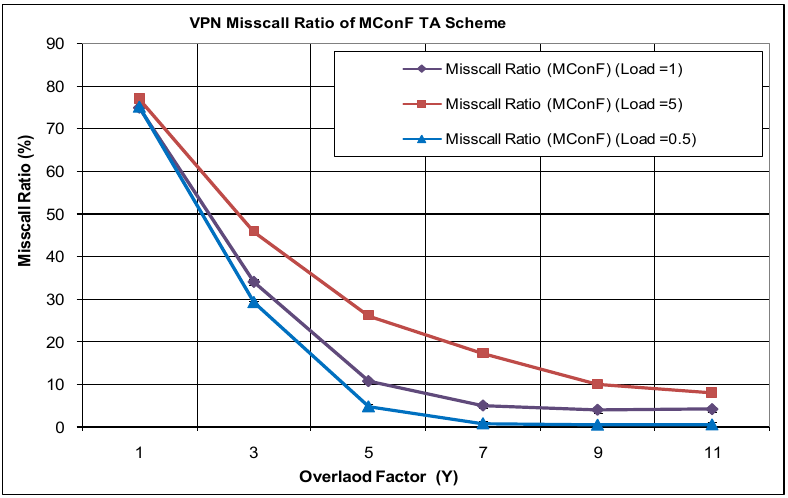} \label{fig:23}}
\subfloat[]{\includegraphics[width=.67\columnwidth, height=45mm]{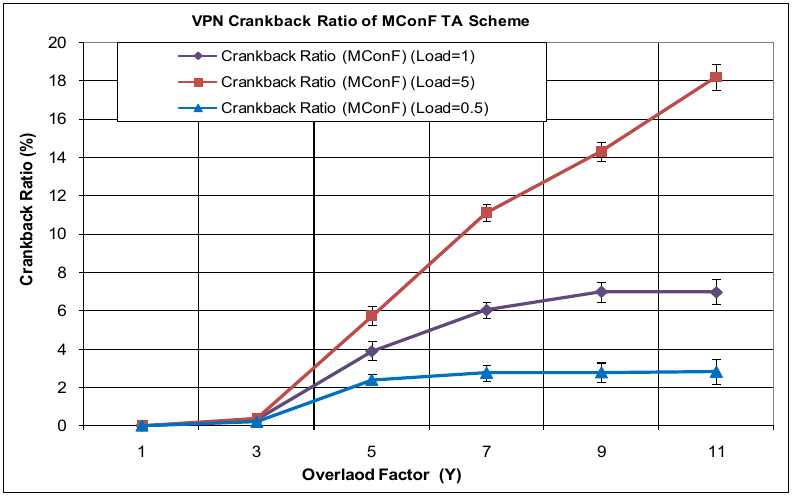} \label{fig:24}}
\subfloat[]{\includegraphics[width=.67\columnwidth, height=45mm]{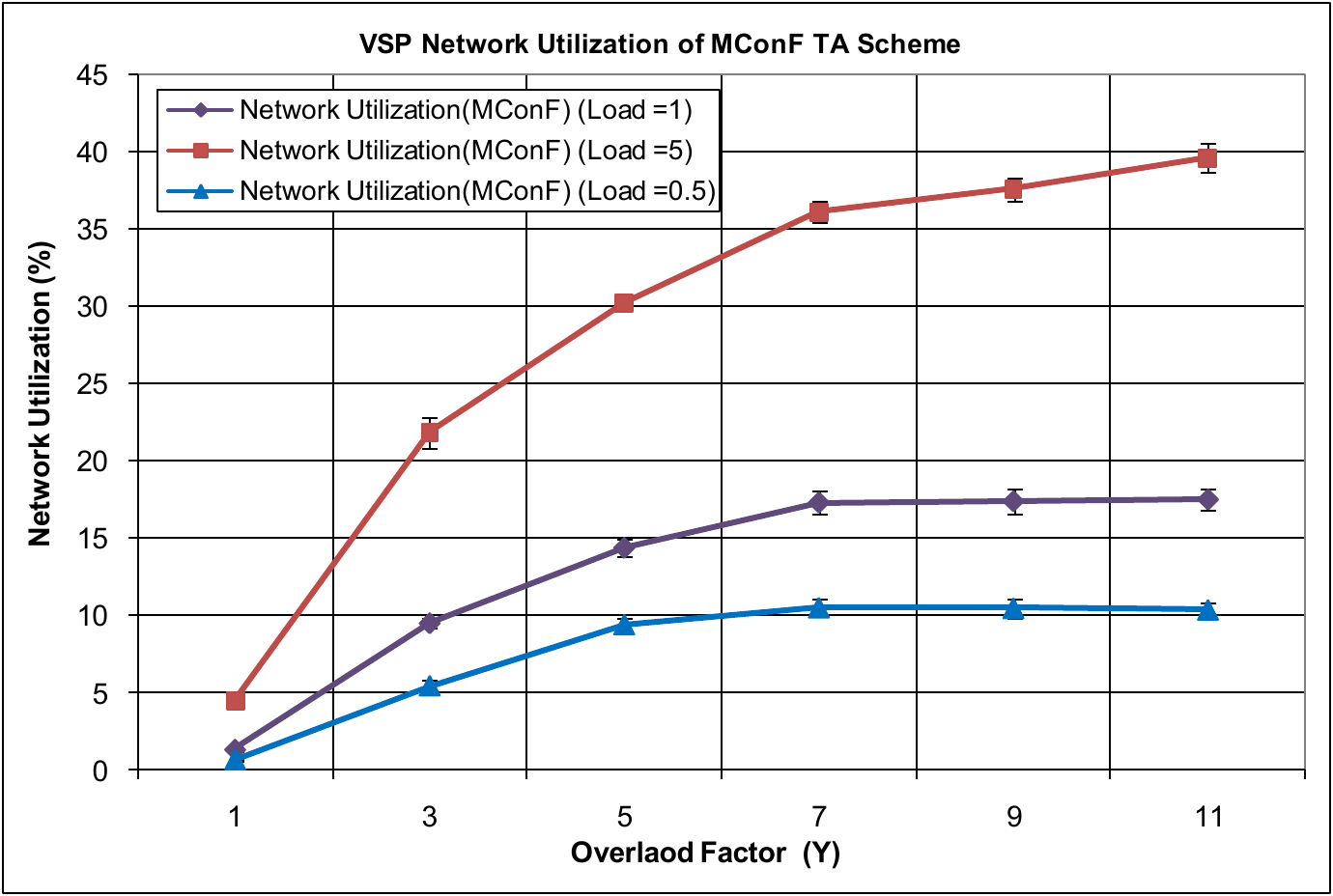} \label{fig:25}}\\
\subfloat[]{\includegraphics[width=.67\columnwidth, height=45mm]{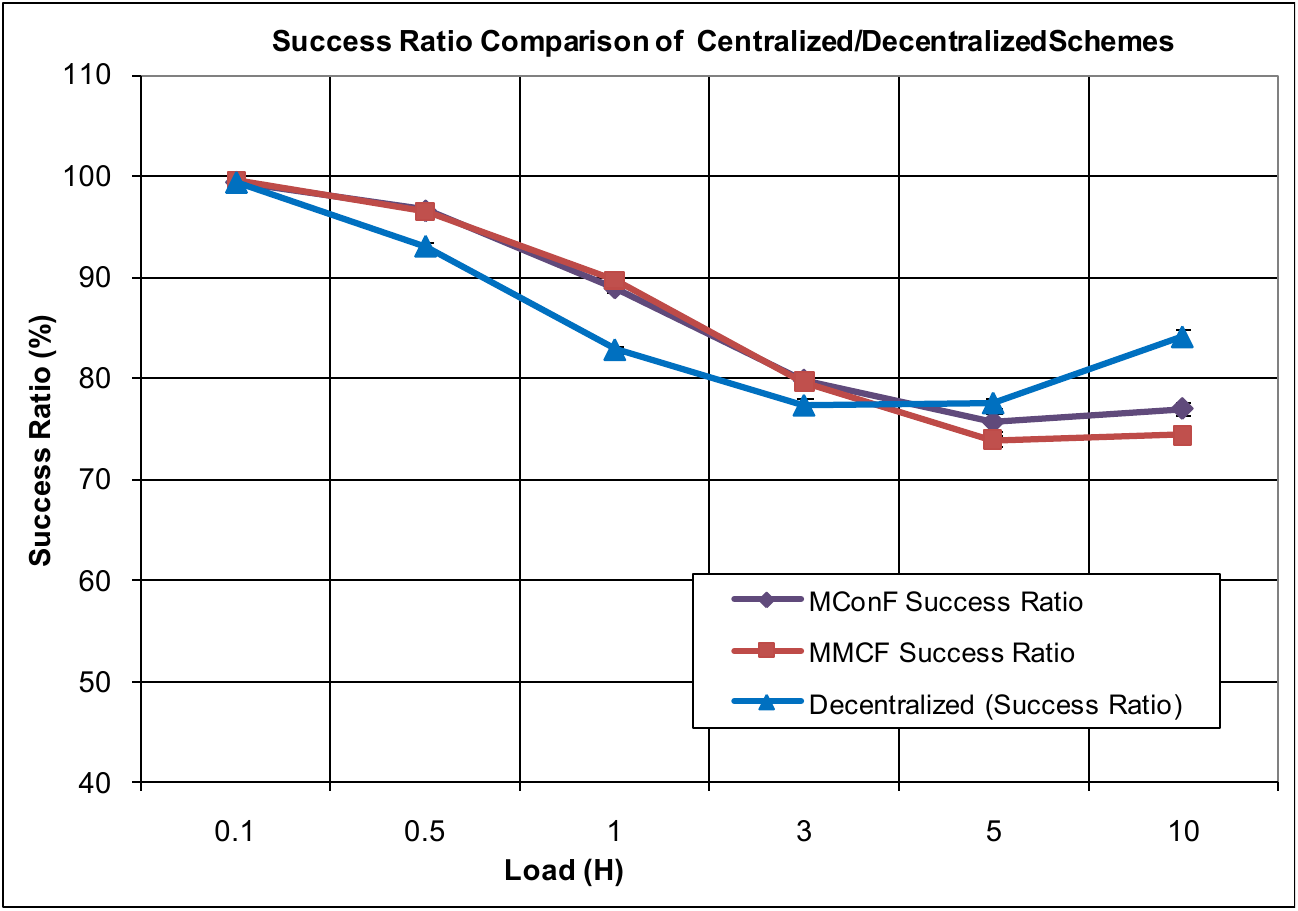} \label{fig:26}}
\subfloat[]{\includegraphics[width=.67\columnwidth, height=45mm]{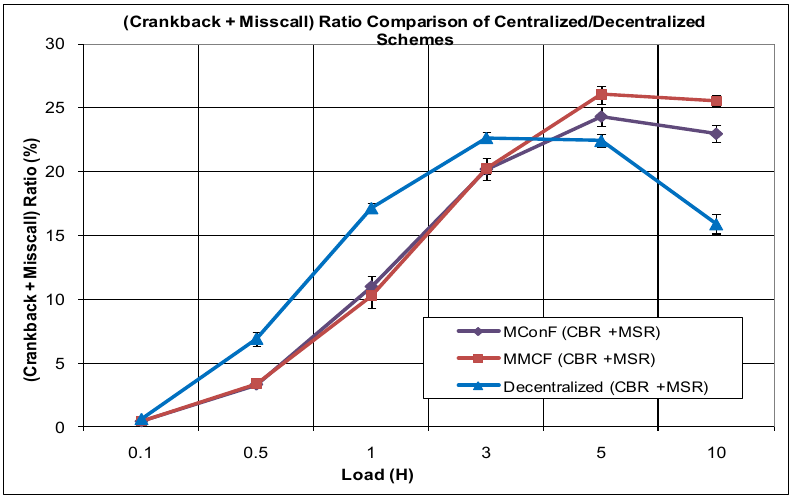} \label{fig:27}}
\caption[]{\small \subref{fig:23} VPN call performance comparing misscall ratio of  MConF based abstraction scheme with varying oversubscription factor (Y); \subref{fig:24} VPN call performance comparing crankback ratio of  MConF based abstraction scheme with varying oversubscription factor (Y); \subref{fig:25} VSP network utilization comparison of  MConF based abstraction scheme with varying oversubscription factor (Y); \subref{fig:26} Success ratio comparison of MConF/MMCF/decentralized TA schemes with varying mean holding time ($\mathcal{H}$); \subref{fig:27} (Crankback +Misscall) ratio comparison of MConF/MMCF/decentralized TA schemes with varying $\mathcal{H}$.}
\vspace{-1.5em}
\end{figure*}

In the next objective we try to improve the misscall ratio performance and in turn the success ratio performance of the centralized schemes by applying the technique of oversubscription of residual link capacities.

{\bf Objective 3(b): Improving the performance of centralized TA schemes by oversubscription of link capacities}

One of the observations from the study of the previous objective is  that oversubscription in the context of the TA service  in a dynamic bandwidth scenario could help improve statistical multiplexing. In this objective, we study if this notion can be used to improve the conservative nature of the centralized schemes, particularly the misscall ratio performance.  We do this by oversubscribing the links of the core network by multiplying the residual capacity of the link by a constant factor (Y)  before executing either MConF or MMCF based subgraph  partitioning schemes. Figures~\ref{fig:21}--\ref{fig:26}
compare the performance of the MConF and MMCF based schemes for a range of values of oversubscription factor Y in the interval [1,10]. The performance statistics are collected for low, medium and high load conditions set to 0.5, 1 and 5 Erlangs.

Figures~\ref{fig:21}--\ref{fig:22}
compares the success ratio with varying values of oversubscription factor for different load conditions for the MConF and MMCF based  schemes. At the low load of 0.5 Erlang,  we can observe that even at Y set to 3, the success ratio of MMCF improves by $50\%$. Also the gain in success ratio is observed to improve with increasing value of $Y$. In Fig.~\ref{fig:23}, 
we see a similar improvement with respect to misscall ratio performance.   From Figs.~\ref{fig:21}--\ref{fig:23}
we also observe that the gain in success and misscall ratio saturates after a particular value of $Y$. For the low load case this value of  $Y$  is $7$. The reason for saturation of the success and misscall ratio performance with increasing oversubscription of  links is due to increase in the crankback ratio (see Fig.~\ref{fig:24})
which achieves its maximum value of $3\%$ when  $Y \ge 7$ for the low load case. With respect to network utilization (see  Fig.~\ref{fig:25}) 
we observe that gain in success ratio also translates into increasing gain in network utilization. The saturation in network utilization is due to the saturation observed with respect to the success ratio metric discussed earlier. Similar trends in gain in success and misscall ratios is observed for network utilization performance for medium and high load cases too.

From Fig.~\ref{fig:26}
we observe that, with increasing load, the overall gain in success ratio from applying oversubscription decreases. That is, the maximum success ratio achieved with the load offered at $0.5$ is $97\%$ compared to $74\%$ when load is $5$. This is due to two reasons. First, this is due to decrease in abstraction efficiency of MConF based partitioning scheme with increasing load as discussed in objective $1$ with reference to Fig.~\ref{fig:11}. 
Second, with increasing load and increasing network utilization, the resource contention among VPNs also increases. This causes the crankback ratio to increase, and this increases with increasing load whose trend can be observed in Fig.~\ref{fig:27}.
Figure~\ref{fig:22}
compares the gain in success ratio achieved for MMCF based abstraction scheme. The trend in performance with increasing oversubscription factor in terms of success ratio in the case of MMCF based abstraction scheme is similar to that observed for MConF based scheme. The important observation here is that the difference in performance between MMCF and MConF based abstraction scheme reduces with increasing link oversubscription. This is because, with increasing oversubscription the advantage gained by better abstraction efficiency of MMCF reduces, hence MConF begins to perform as good as MMCF at higher oversubscription factors. The reason for decreasing trend of the success ratio performance with increasing load for the MMCF case is the same as those for the MConF case.

Next we compare the performance of decentralized scheme with MConF and MMCF schemes after link oversubscription is applied. For this scenario, based on the previous results we choose an oversubscription factor of 9. Figure~\ref{fig:26}
compares the success ratio for the three schemes with link oversubscription. Here we observe that MConF and MMCF perform better than the decentralized scheme particularly at low and medium load conditions. We see that MConF and MMCF schemes perform on an average $4\%$ better than the decentralized scheme. We also observe that the difference between performance of MConF and MMCF itself is not significant. This is because, the link oversubscription overshadows the gain in abstraction efficiency achieved by MMCF  which was discussed earlier. At high loads, we see that the decentralized scheme performs better than MConF and MMCF based TA scheme. This is because at high loads the crankback and misscall ratios, as shown in Fig.~\ref{fig:27}, 
deteriorates for MConF and MMCF based schemes, resulting in poorer success ratio than the decentralized scheme. This shows that oversubscription may not always act favorably in all load conditions. This is so because link oversubscription at high load conditions results in abstractions that are too aggressive causing VPNs to make poor decisions resulting in high crankback and misscall ratios. In such situations the VSP could adopt the technique of using a tunable oversubscription factor that varies with utilization of network resources.

To summarize the above observations , we note that oversubscribing the core links by a fixed factor does help to overcome the conservative nature of centralized MConF and MMCF based schemes  to a significant extent.
\section{Discussion}
\label{sec:sec7}

We would like to conclude by pointing out opportunities for extending our works presented in this paper.

Given a graph $G$ with vertex set $V$ and a VPN  defined on a subset $V^\prime$ of $V$, a steiner logical topology (or graph) $G^\prime$  is a graph defined on $V^\prime$ such that $G^\prime$ provides an accurate estimate of the maximum flow available in $G$ between the vertices of $G^\prime$. The steiner logical topology problem is to design a steiner logical topology for a given subset $V^\prime$ of $V$.  We note that a steiner logical graph is not necessarily a subgraph of $G$. So we need to assign capacities to the links in the steiner logical graph to achieve the desired property.  So, there are many ways to do this. For instance, we can assign each logical link with a path in $G$ (similar to lightpath in IP-over-WDM optical networks). What makes the problem challenging is the fact that several VPNs will be present simultaneously on a given $G$ and all of them need to be provided with an accurate estimate of the available capacities in $G$.  Certain recent works and the references therein can provide the basis for further research in this direction~\cite{peleg1989}--\cite{kt2010}.

Another area where TA can be used is to address the problems arising from conflicting traffic engineering (TE) principles applied by over-the-top (OTT) content providers and the underlay service providers~\cite{keralapura2005}. This problem could result in unstable network behavior. This situation can be improved by adopting a cooperative model between the OTT and the SP, wherein the SP could expose the underlying physical network's properties in the form of a TA to the OTT content provider. This form of cooperative benefit has been studied in~\cite{jiang2007}, but in the context of the provider providing the complete topology to the OTT provider which is not feasible practically. TA could be a promising direction to address this problem.
\section{Conclusion}\label{sec:conclusion}
This paper is a continuation of~\cite{ravi2006}\cite{ravi2013} where we introduced the TA service for VPNs and proposed three decentralized schemes. These schemes assume that all the border nodes performing the abstraction have access to the entire core network topology.  In contrast to this, in this paper we have developed centralized schemes to partition the core network capacities, and assign each partition to a specific VPN for applying the decentralized abstraction schemes presented in~\cite{ravi2006}\cite{ravi2013}.  Towards this end, we first introduced the VPN core capacity sharing (VPN-CS) problem. We applied multicommodity flow theory to solve this problem. Considering the objectives of the VPN-CS problem, we first proposed a method based on the maximum concurrent flow theory, and later improved it by proposing the use of the maximum multicommodity flow theory. In order to address the fairness issue associated with the MMCF based approach, a new problem called the fair partitioning problem was formulated. We studied this problem by proposing the bounded MMCF formulation, wherein two variations were proposed. In addition, we also proposed a heuristic called the flow balancing scheme to address the fair partitioning problem for online implementation.  Performance analysis was conducted where we compared the proposed schemes using offline implementations with an LP tool and discrete event driven simulation.

From the simulation analysis, we observed that the MMCF based partitioning approach resulted in better call performance than the MConF based partitioning scheme. The MMCF based partitioning algorithm performed about $5\%$ better than the MConF approach with respect to both VPN call performance and network utilization metrics. With respect to abstraction efficiency, we observed that MMCF's performance was $37\%$ better than MConF; however, in terms of fairness, the standard deviation of the commodity flows generated by MConF was on an average $75\%$ less than the standard deviation of the flows generated by MMCF. With flow balancing enabled, the simulation analysis also showed that the fairness issue of MMCF could be improved by about $25\%$ with no significant impact on the overall call performance of the VPNs or the core network utilization.

We also studied how the conservative nature of the centralized schemes can be overcome by oversubscribing the residual capacity of core network links before applying the centralized abstraction schemes. We observed that the call performance and core network utilization showed significant improvement over a range of oversubscription factors, demonstrating this as a practical way to overcome the conservative nature of the centralized mode of VPN TA generation.

Another line of investigation is to study how best one can use the ideas of partitioning developed in this paper  for enhancing the capabilities of the Virtual Cluster Embedding schemes discussed in \cite{chowdhury12}\cite{rost15} and the references therein.

\begin{IEEEbiography}[{\includegraphics[width=1in,height=1.25in,clip,keepaspectratio]{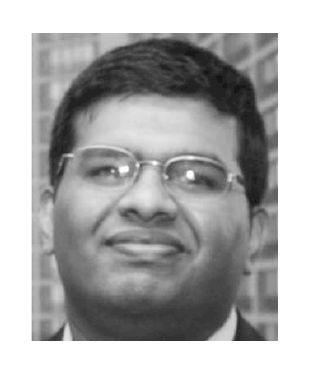}}]{Ravishankar~Ravindran}
 received the PhD degree in electrical engineering from Carleton University, Ottawa, Canada and the MS degree in computer science from the University of Oklahoma, Norman. He is currently a senior researcher at Huawei's Research Center, Santa Clara, conducting research in area of future Internet architectures. Before this, he was a part of Nortel's Advanced Technology group where he conducted research in the areas of Optical Networking, QoS Routing, Control Plane Architectures related to IP/(G)MPLS, 4G Wireless Research, and End-to-End QoE/QoS Engineering for Multimedia Applications. He is a senior member of the IEEE.
\end{IEEEbiography}
\begin{IEEEbiography}[{\includegraphics[width=1in,height=1.25in,clip,keepaspectratio]{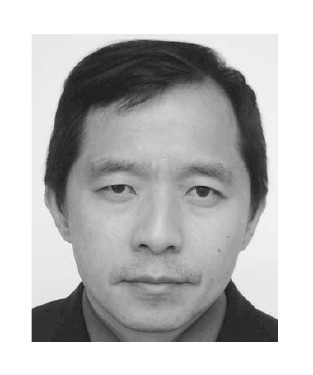}}]{Changcheng~Huang} received the BEng degree in 1985 and the MEng degree in 1988 both in electronic engineering from Tsinghua University, Beijing, China. He received the PhD degree in electrical engineering from Carleton University, Ottawa, Canada (1997). From 1996 to 1998, he worked for Nortel Networks, Ottawa, Canada where he was a systems engineering specialist. He was a systems engineer and network architect in the Optical Networking Group of Tellabs, Illinois, 1998 -- 2000. Since July 2000, he has been with the Department of Systems and Computer Engineering at Carleton University, Ottawa, Canada and is now a full professor. He won the CFI new opportunity award for building an optical network laboratory in 2001. He was an associate editor of IEEE Communications Letters from 2004 to 2006. He is a senior member of the IEEE.
\end{IEEEbiography}
\begin{IEEEbiography}[{\includegraphics[width=1in,height=1.25in,clip,keepaspectratio]{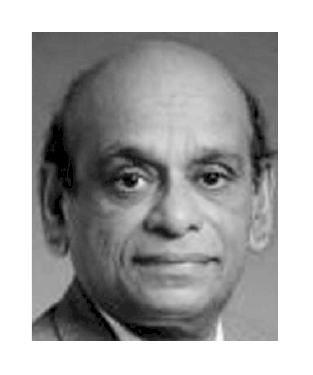}}]{Krishnaiyan~Thulasiraman}
 received the PhD degree in EE from IIT Madras, India, in 1968 and has been a professor and holds the Hitachi chair in CS at the University of Oklahoma since 1994.
His research has been in graph theory, combinatorial optimization, algorithms, and applications in a variety of areas in CS and EE. He has coauthored with M.N.S. Swamy two text books Graphs, Networks, and Algorithms (1981) and Graphs: Theory and Algorithms (1992), both published by Wiley Inter-Science. He has received several awards and honors: Distinguished Alumnus Award of IIT Madras(2008), Fellow of the American Association for Advancement of Science (2007), 2006 IEEE Circuits and Systems Society Technical Achievement Award.
\end{IEEEbiography}
\begin{IEEEbiography}[{\includegraphics[width=1in,height=1.25in,clip,keepaspectratio]{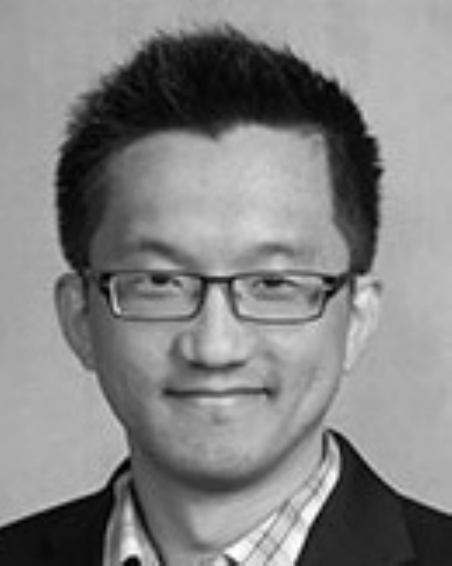}}]{Tachun~Lin} received the Ph.D. degree in Computer Science from the University of Oklahoma, and the M.S. and B.S. degrees from the National Chiao Tung University, Hsinchu, Taiwan. He is currently an Assistant Professor at the Department of Computer Science \& Information Systems, Bradley University. His research interests are in cross-layer interdependent networks, network optimization, mathematical programming, graph theory, and game theory. He is a member of the ACM, IEEE, and INFORMS.
\end{IEEEbiography}

\vfill

\end{document}